\documentclass[reprint,superscriptaddress,amsmath,amssymb,amsfonts,aps,prb,floatfix]{revtex4-2}
\usepackage{times}
\usepackage{graphicx}
\usepackage{dcolumn}
\usepackage{bm}
\usepackage[colorlinks=true, citecolor=blue, linkcolor=blue, urlcolor=blue]{hyperref}
\usepackage{bbold}
\DeclareMathAlphabet\mathbfcal{OMS}{cmsy}{b}{n}
\usepackage{wasysym}
\usepackage{amssymb}
\usepackage{amsmath}
\usepackage[usenames,dvipsnames]{xcolor}

\begin{document}
\title{Ephemeral Superconductivity Atop the False Vacuum}

\author{Gal Shavit}
\affiliation{Department of Physics and Institute for Quantum Information and Matter, California Institute of Technology,
Pasadena, California 91125, USA}
\affiliation{Walter Burke Institute of Theoretical Physics, California Institute of Technology, Pasadena, California 91125, USA}
\author{Stevan Nadj-Perge}
\affiliation{Department of Physics and Institute for Quantum Information and Matter, California Institute of Technology,
Pasadena, California 91125, USA}
\affiliation{T. J. Watson Laboratory of Applied Physics, California Institute of
  Technology, 1200 East California Boulevard, Pasadena, California 91125, USA}
\author{Gil Refael}
\affiliation{Department of Physics and Institute for Quantum Information and Matter, California Institute of Technology,
Pasadena, California 91125, USA}

\maketitle

\section*{Abstract}
A many-body system in the vicinity of a first-order phase transition may get trapped in a local minimum of the free energy landscape.
These so-called false-vacuum states may survive for exceedingly long times if the barrier for their decay is high enough.
The rich phase diagram obtained in graphene multilayer devices presents a unique opportunity to explore transient superconductivity on top of a correlated false vacuum.
Specifically, we consider superconductors which are terminated by an apparent first-order phase transition to a correlated phase with different symmetry.
We propose that quenching across this transition leads to a non-equilibrium ephemeral superconductor, readily detectable using straightforward transport measurements.
Moreover, the transient superconductor also generically enhances the false vacuum lifetime, potentially by orders of magnitude.
In several scenarios, the complimentary effect takes place as well: superconductivity is temporarily emboldened in the false vacuum, albeit ultimately decaying.
We demonstrate the applicability of these claims for different instances of superconductivity terminated by a first order transition in rhombohedral graphene.
The obtained decay timescales position this class of materials as a promising playground to unambiguously realize and measure non-equilibrium superconductivity.

\section*{Introduction}

\begin{figure}
    \centering
    \includegraphics[width=8.5cm]{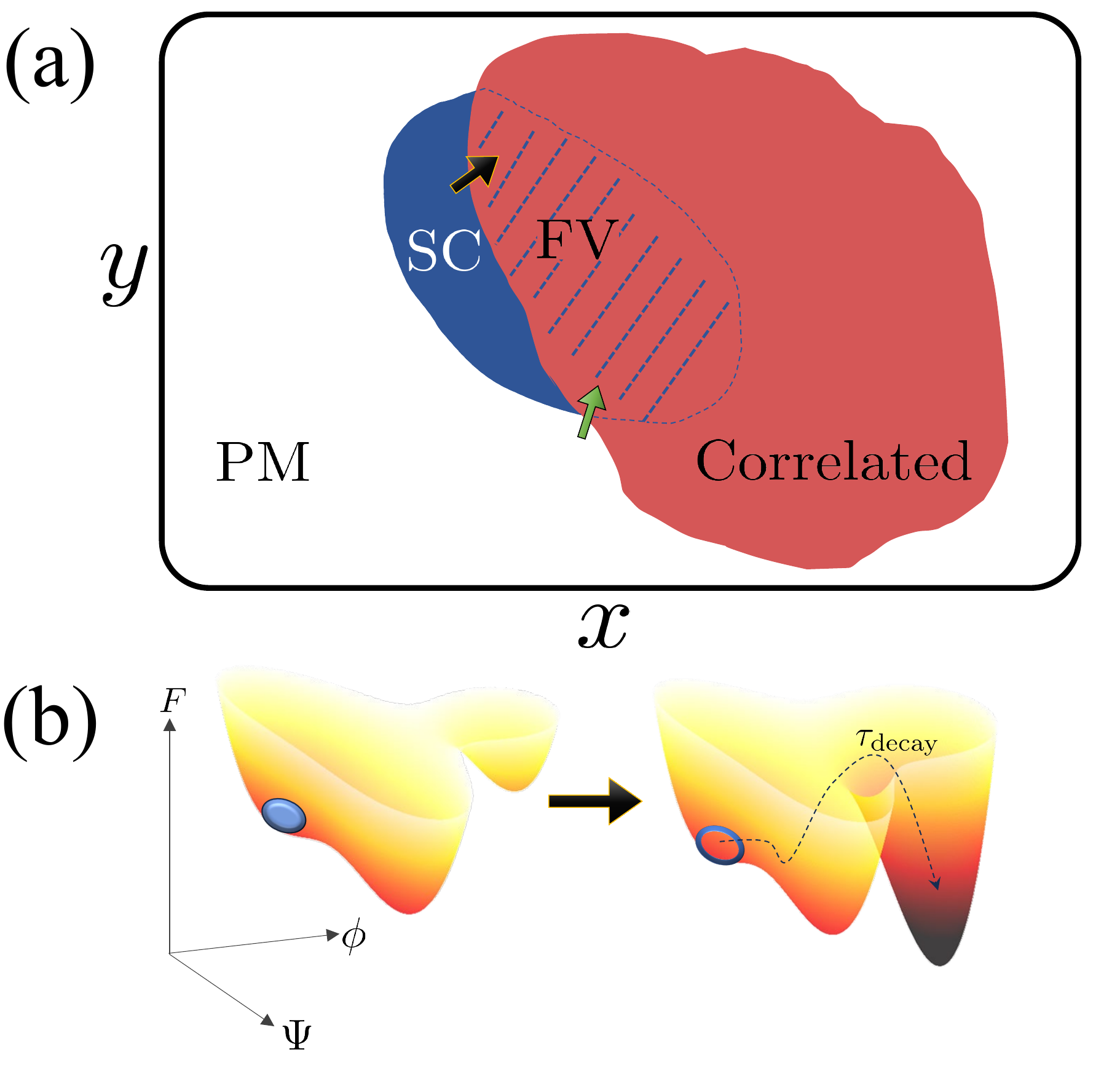}
    \caption{
    \textbf{Schematic equilibrium phase diagram hosting ephemeral superconductivity.}
    (a)
    As a function of tuning parameters ($x,y$) a superconducting region exists (SC, blue), surrounded by a parent ``normal'' state (PM, white) and a symmetry-broken correlated phase (red).
    The correlated phase suppresses an otherwise-superconducting region in parameter space (hatched red-blue).
    Here, we consider a quench across a first-order transition boundary to the FV superconductor, either directly from an equilibrium superconducting phase (black arrow), or from the normal state (green arrow).
    (b)
    Illustration of the energy landscape as a function of the two order parameters $\phi,\Psi$ across the transition.
    In the equilibrium superconducting phase (left) $\Psi$ is condensed (blue ellipse).
    Following the quench (black arrow), the new equilibrium state has $\phi$ condensation and zero superconductivity.
    The ephemeral superconductor (empty ellipse) and has a finite lifetime $\sim \tau_{\rm decay}$ before decaying to the true vacuum (via the dashed line trajectory).
    }
    \label{fig:FVphasediagram}
\end{figure}

Phase transitions in many-body correlated systems may often be succinctly described by an appropriate classical or quantum field theory~\cite{altland2010condensed,wen2004quantumBook}.
The equilibrium many-body ground state is identified by the global minimum of the free-energy associated with this description.
However, other locally-stable minima may exist, albeit with a higher energy density.

These minima act as the ``false-vacuum'' (FV) of the system, and may be long-lived due to their metastable nature.
A system can generically get trapped in the FV state when it is quenched through a first-order phase transition.
Supercooling and superheating of water are well-known classical examples of this phenomenon~\cite{debenedetti2020metastableLiquids}, yet quantum systems such as spin chains~\cite{analyticIsing,FVspinchains,PRXQuantum.3.020316FVspins}, superconducting wires~\cite{MccumberHalperin}, and atomic superfluids~\cite{FVatomicsuperfluids} 
have been show to exhibit metastable phases and FV decay.
Further, the FV concept itself has originated in the context of cosmology, where it may have truly dire implications~\cite{Coleman_SemiclassicalPhysRevD.15.2929,Coleman_Quantum_PhysRevD.16.1762,Turner1982Wilczek}.

In recent years, graphene multilayers have emerged as an exciting platform with a high degree of tunability to study correlated electron phenomena, topological phases, unusual superconductivity, and their interplay~\cite{TBG1_CaoCorrelatedInsulator,TBG_2CaoUnconventionalSC,TBG3_EfetovAllIntegers,KimTrilayerGrapheneSuperconductivity,Park2021StronglyCoupledSuperconductivityTrilayer,Park2022MATngFamily,matngYIRAN,TDBG1_Shen2020,TDBG2_Cao2020,TDBG3_Liu2020,TDBG_Hysteresis_folkKuiri2022,TDBG_folk_Su2023,Quasicrystal_Uri2023,RTG_half_quarter_metal_Zhou2021,RTGsuperconductivityZhou2021,ZhouYoungBLGZeeman,nadj_ISOC_BBGBLGZhang2023,BBGcorrelated_Seiler2022,BBG_electronside_li2024tunable,PentaG1_Han2023,PentaG2_Han2024,TetraG_Liu2024,PENTA_FCILu2024}.
A recurring theme in these systems is the peculiar vicinity of the superconducting phases to symmetry-breaking phase transitions.
In several cases, the superconducting dome itself is terminated by an abrupt transition where the Fermi surface undergoes significant reconstruction~\cite{ZhouYoungBLGZeeman,RTGsuperconductivityZhou2021,nadj_ISOC_BBGBLGZhang2023,TDBG_folk_Su2023,Quasicrystal_Uri2023,BBG_electronside_li2024tunable,LongJuyang2024diverseimpactsspinorbitcoupling,Nadj_zhang2024twistprogrammablesuperconductivityspinorbitcoupled}, which in some instances is strongly indicative of a first-order phase transition~\cite{young_nadj_BBG_RTG_SC,Young_patterson2024superconductivityspincantingspinorbit,Young_choi2024electricfieldcontrolsuperconductivity}.
Recently, similar phenomenology was observed in twisted bilayer WSe$_2$, where hysteretic behavior was observed at the boundary between superconductivity an a correlated phase~\cite{Wse2_hysteresis_CoryDean_guo2024superconductivity}.

In this work, we propose these materials as a platform for realizing and exploring out-of-equilibrium superconductivity, which exists as a metastable phase on top of the FV manifold of the symmetry-broken phase.

This extraordinary non-equilibrium metastable state arises in the vicinity of the true vacuum symmetry-broken phase.
A useful heuristic of the sort of scenarios we discuss is presented in Fig.~\ref{fig:FVphasediagram}.
In this generic phase diagram, superconductivity and a correlated phase are in close proximity, separated by a first-order transition line.
Specifically, we are interested in cases where the correlated phase preempts superconductivity and 
overtakes it.
Thus, after a sudden quench from the superconducting phase across the transition (black arrow in Fig.~\ref{fig:FVphasediagram}), there exists a possibility of a long-lived transient superconductor, realized on top of the FV.
Alternatively, we also consider quenching from a parent normal phase directly into the suppressed superconductor through a first-order transition (green arrow).
Such a protocol may allow one to 
reveal buried underlying
superconductivity in such systems, masked by competing phases.

We underline the regimes where FV superconductivity are most relevant and experimentally accessible.
This is accomplished by combining microscopic calculations for two candidate materials, rhombohedral trilayer graphene (RTG) and Bernal-stacked bilayer graphene (BBG), and a field theoretical description of the FV decay phenomena.
We estimate the expected lifetimes of the ephemeral superconductors to be of the order of $\sim 100$ nanoseconds, enabling straightforward detection methods, relying on time-resolved transport measurements.
Remarkably, the unusual presence of superconductivity in the FV state is what \textit{enables} such simplified detection schemes in a solid-state setting.
Superconductivity provides an unambiguous transport signal -- a delay between a current driven through the system and the appearance of a voltage drop.

Furthermore, we show that the incompatibility between the correlated symmetry-broken phase and the superconductor \textit{non-trivially enhances the stability of the FV and its lifetime}.
As we show, this is a generic feature in scenarios where a subordinate phase develops on top of a ``primary'' false vacuum.
This may be understood as a result of magnification of the surface tension between the true and false vacuum states of the system.
The strength of surface tension plays a major role in determining the energetics of the FV decay.

Interestingly, the FV superconductivity may actually survive at higher temperatures compared to its equilibrium counterpart on the other side of the transition.
This transient enhancement of superconductivity comes at a cost of incurring a finite lifetime.
Generically, what drives the symmetry-breaking transition which terminates the superconductor is the density of states (DOS) near the Fermi level $\nu$.
Clearly, it is also an important factor in determination of the superconducting properties.
For example, conventionally the superconducting transition temperature $T_c\propto \exp\left(-\frac{1}{u\nu}\right)$ ($u$ is the pairing strength).
For weak-coupling superconductors, $u\nu\ll 1$, $T_c$ is especially sensitive to $\nu$.
In the FV, the superconductor temporarily experiences a higher DOS while the correlated phase is suppressed.
This facilitates favorable superconducting properties, potentially beyond those available at equilibrium under similar conditions. 


\section*{Results}
\subsection*{False vacuum decay}\label{sec:generalframework}

We consider scenarios where as a function of some tuning parameter, $r$ (which can be a magnetic field, an electric displacement field, pressure, etc.), a system undergoes a first-order phase transition with an order parameter $\hat{\phi}$.
Within this $\hat{\phi}$-ordered phase, it is further assumed that the disordered (e.g., paramagnetic) phase
remains a metastable local minimum of the effective free energy.

Let us also examine the consequences of an additional phase, with order parameter $\hat{\Psi}$, which (i) also condenses in the vicinity of the $\hat{\phi}$ phase transition, and (ii) is \textit{incompatible} with the $\hat{\phi}$ phase.
Generally, we will be interested in the case where the energy scales associated with $\hat{\phi}$ dominate over those of $\hat{\Psi}$, and the $\hat{\phi}$ order is the equilibrium ground state, and the so-called ``true vacuum''.

The scenario we describe above may be captured by the following Ginzburg-Landau free energy functional, 
\begin{align}
    F	&=\int d^{2}x\left[\frac{\sigma}{2}\left|\nabla\phi\right|^{2}+16g\phi^2\left(\phi-1\right)^{2}-B\phi^2\right]\nonumber\\
	&+\int d^{2}x\left[\frac{\kappa}{2}\left|\nabla\Psi\right|^{2}-\frac{a\left(r\right)}{2}\left|\Psi\right|^{2}+\frac{b}{4}\left|\Psi\right|^{4}\right] \nonumber\\
	&+\frac{\lambda}{2}\int d^{2}x\left|\Psi\right|^{2}\phi^{2}. \label{eq:GLasymm}
\end{align}
Clearly, the first-order transition is governed by the parameter denoted as $B$.
When $B<0$, $\phi=0$ (the paramagnetic phase where the spontaneous ordering of $\hat \phi$ does not yet occur)
minimizes the free energy in the absence of $\hat{\Psi}$.
At $B=0$, the minima at $\phi=0\,, 1$ are degenerate, and the ordered phase takes over at $B>0$, where $\phi\approx 1$ becomes the global minimum.
In the metastable regime $B<B_c=16g$, there exists a finite energy barrier between the two minima, whose strength is $g$ right at the transition.
The stiffness $\sigma$ gives rise to a surface tension between domains of $\phi=0$ and $\phi\approx 1$,
\begin{equation}
    J_\phi = \int_0^{\phi_{>0}} d\phi\sqrt{2\sigma V\left(\phi\right)},
    \label{eq:surfacetensionInstanton}
\end{equation}
where $V\left(\phi\right)=16g\phi^{2}\left(\phi-1\right)^{2}-B\phi^2$, and $\phi_{>0}$ is the maximal $\phi$ for which $V\left(\phi\right)>0$.
The correlation length is calculated by optimizing the energy of a domain-wall for $B=0$ (See Supplementary Section S1), $\xi_\phi=\sqrt{\frac{\sigma}{8g}}$.

In the second part of $F$, $a\left(r\right)$ is associated with the second-order transition.
When $a>0$, the decoupled uniform solution is $\left|\Psi\right|=\Psi_0\sqrt{\frac{a}{b}}$.
Here the appropriate correlation (coherence) length is $\xi_\Psi=\sqrt{\kappa/\left(2a\right)}$, and the surface tension is 
$J_\Psi = \frac{8}{3}\xi_\Psi\frac{a^{2}}{4b}$.
Finally, $\lambda>0$ couples the two sectors, and precludes $\hat{\phi}$-$\hat{\Psi}$ order coexistence when it is sufficiently large as compared to the energy densities $g$ and $a$.

\begin{figure}
    \centering
    \includegraphics[width=9cm]{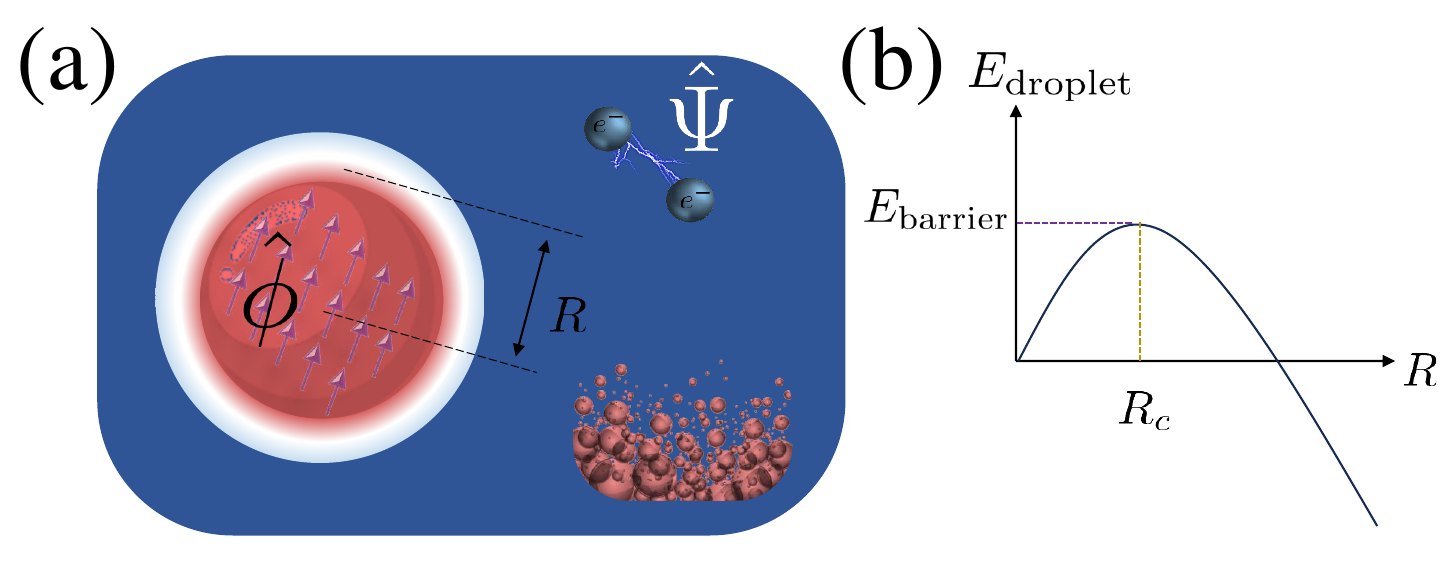}
    \caption{
    \textbf{Critical droplet formation and analysis.}
    (a)
    Illustration of a bubble of the true vacuum phase $\hat{\phi}$ (e.g., ferromagnet) within a metastable bulk $\hat{\Psi}$ phase (superconductor).
    Due to the incompatibility of the phases, there exists a narrow ``corona'' where $\hat{\Psi}$ is first suppressed, and only then $\hat{\phi}$ is established.
    (b)
    Schematic dependence of the bubble energy on its radius.
    Due to surface tension contributions, the droplet energy first increases with $R$, before eventually trending towards expansion.
    The critical radius $R_c$, as well as the threshold energy $E_{\rm thresh.}$, are indicated.
    }
    \label{fig:bubblefigure}
\end{figure}

We consider the system described by Eq.~\eqref{eq:GLasymm}, initialized to the ordered (superconducting) $\Psi_0$ phase, yet tuned to metastability, $a>0$ and $0<B<B_c$.
The true vacuum is $\phi\approx1$, yet the initial state of the system is a locally stable minimum of the free energy.
We schematically evaluate the characteristic time necessary for the system to decay to the true vacuum, adapting the methods developed by Langer~\cite{LANGER1967108} and Coleman~\cite{Coleman_SemiclassicalPhysRevD.15.2929}.

Neglecting quantum fluctuations (whose role we later discuss), 
relaxation is facilitated by thermal fluctuations of $\hat{\phi}$ droplets (or ``bounces''~\cite{altland2010condensed}) within the $\hat{\Psi}$ bulk, see illustration in Fig.~\ref{fig:bubblefigure}a.
The energy of a droplet with radius $R$ may be evaluated in the so-called thin-wall limit ($R\gg \xi_\phi$) by~\cite{Coleman_SemiclassicalPhysRevD.15.2929,altland2010condensed} 
\begin{align}
    E_{\rm droplet}&=\pi R^2 f_{\rm true}
    -\pi \left(R+\delta R\right)^2  f_{\rm meta}\nonumber\\
    &+2\pi R J_\phi + 2\pi \left(R+\delta R\right) J_\Psi, \label{eq:Ebubblecomposite}
\end{align}
where $f_{\rm true}=-B$, $f_{\rm meta}=-\frac{a^2}{4b}$, and $\delta R \approx \left(\xi_{\phi}+\xi_{\Psi}\right)/2$, due to the necessary suppression of the $\hat{\Psi}$ preceding the $\hat{\phi}$ droplet formation.
The droplet experiences an effective force $\propto \partial E_{\rm droplet}/\partial R$, pushing it towards either expansion or collapse, thus determining its fate.
Thus, the droplet energy threshold, i.e., the droplet energy at which it would tend to overtake the system, will be given by 
\begin{equation}
    E_{\rm thresh.}=E_{\rm droplet}\left(R_c\right),
\end{equation} 
and
\begin{equation}
    R_{c}=\frac{J_{\phi}+J_{\Psi}-\delta Rf_{{\rm meta}}}{f_{{\rm meta}}-f_{{\rm true}}},\label{eq:Rc} 
\end{equation}
determined by $\partial E_{\rm droplet}/\partial R |_{R_c}=0$ (Fig.~\ref{fig:bubblefigure}b).
The expression for $R_c$ accounts for two effects, which are \textit{solely due to the ordered $\hat{\Psi}$ phase}, on top of the FV.

First, the denominator in Eq.~\eqref{eq:Rc} is made smaller  by the presence of the finite condensation energy of $\hat{\Psi}$.
In our regime of interest, as we discuss below, one usually finds $\left|f_{\rm true}\right|\gg\left|f_{\rm meta}\right|$, and this effect is of vanishing importance.

Second, the additional $J_\Psi$ in the numerator, associated with the surface tension of the secondary (superconducting) phase, is reasonably expected to be rather small compared to $J_\phi$, the surface tension contribution of the dominant correlated order, yet not negligible.
This is because the surface tension is roughly proportional to the product of the correlation length and the energy scale associated with the ordered phase.
While the latter might be much bigger for $\hat{\phi}$ as compared to $\hat{\Psi}$, this difference is usually somewhat compensated by the ration of correlation lengths, $\xi_\Psi\gg\xi_\phi$.
In the regime where the denominator effect is negligible, one may approximate the threshold droplet radius
$R_c\approx J_\phi/B \left(1+c J_\Psi/J_\phi\right)$, where $c$ is an order-1 numerical factor which depends on microscopic details.
As we demonstrate for the particular systems we consider below, this mechanism indeed facilitates a more stable false-vacuum in the presence of the secondary $\hat{\Psi}$ phase, which in our case is a superconductor.

Our analysis utilizes the classical Ginzburg-Landau functional, Eq.~\eqref{eq:GLasymm}, producing a free-energy barrier towards nucleation and eventual FV decay.
Let us discuss the role of quantum fluctuations on the phenomenon described here.
In the zero temperature limit, $\beta=\frac{1}{T}\to \infty$, any finite energy barrier completely annihilates decay through thermal fluctuations.
However, fluctuations due to the quantum nature of the $\hat{\phi}$ field may overcome this limitation.

Setting aside $\hat{\Psi}$ for a moment, one interprets the $\hat{\phi}$ part of Eq.~\eqref{eq:GLasymm} as the classical limit of the imaginary-time action
$S_\phi=\int_0^\beta d\tau \int d^{d}x 
\left[\frac{\rho}{2}\left(\partial_{\tau}\phi\right)^{2}+\frac{\sigma}{2}\left|\nabla\phi\right|^{2}+V\left(\phi\right)\right]$.
This corresponds to the imaginary-time path integral partition function
$Z=\int D\phi e^{-S_\phi}$.
In the $\rho\to\infty$ limit, variations along the $\tau$ dimension are suppressed, and one recovers the classical limit (with a prefactor of $\beta$ replacing $\int d\tau$ integration).

Performing similar nucleation calculus in Euclidean space-time (see Supplementary Section S1), one finds a temperature scale, $T_Q$, below which the quantum decay pathway is dominant,
\begin{equation}
    T_Q=\frac{3}{32}\frac{B\xi_{\phi}}{J_{\phi}}\tau_{\phi}^{-1},
\end{equation}
with $\tau_{\phi}=\sqrt{\frac{\rho}{8g}}$.
We stress that below $T_Q$ the FV phenomenon persists, yet the lifespan of the metastable phase \textit{saturates}, and remains roughly the same as the temperature is lowered further.

Notably, at any finite temperature, a small enough bias $B$ exists such that the quantum decay is much less efficient.
The reason for this dependence on the bias $B$ stems from the effective higher dimension of the quantum problem, where the surface of the droplet is $d$-dimensional, whereas classically it is $\left(d-1\right)$-dimensional.
As a consequence (see Supplementary Section S1), the threshold energy scales as $B^{-d}$ in the quantum case ($B^{1-d}$ classically).
Thus, when the bias between the false and true vacuum is small enough, the quantum decay pathway is highly disfavored.
Without loss of generality, we henceforth assume this is the case for our detailed microscopic analysis below.

Another interesting possibility occurs in an intermediate temperature regime, where quantum behavior dominates the $\hat{\phi}$ sector, yet $\hat{\Psi}$  (now introduced back in our discussion), remains effectively classical due to its much longer correlation time $\tau_\Psi$.
The relevant temperature regime is
$\frac{\tau_\phi}{\tau_\Psi}<T/T_Q<1$.
Heuristically, it takes considerably lower temperatures (as compared to $T_Q$) to saturate the effects of the $\hat{\Psi}$ field redistribution energetics.

This regime thus enables further enhancement of the FV stability by the competing sub-leading order, as discussed in Supplementary Section S1.
Once more, this enhancement takes place even if the discrepancy between 
$f_{\rm true}$
and
$f_{\rm meta}$
is of orders of magnitude, due to combination of the correlation length effect $\xi_\Psi\gg\xi_\phi$ (already discussed above), and a $\propto T_Q/T$ prefactor to the relative $\hat{\Psi}$ contribution.
Thus, the lower the temperature in this regime, the larger the FV life-time enhancement becomes due to the ordering competition.

\subsection*{Realization in graphene multilayers}\label{sec:grapheneapplications}

We consider first-order phase boundaries observed in both BBG and RTG separating correlated symmetry-broken phases and superconductivity, either incipient or fully formed.
In all cases, we follow the same methodology allowing us to extract the relevant energy densities and surface tensions.

We begin with an Hamiltonian of the form
 \begin{equation}
     H= H_0 + H_{\rm int},
 \end{equation}
where $H_0$ describes the appropriate band structure, and 
$H_{\rm int}=\frac{U_c}{\Omega}\sum_{\bf q} \rho_{\bf q} \rho_{\bf -q}$ 
is the electron-electron interaction, which we take as short-range for simplicity ($\rho_{\bf q}$ is the momentum-$\bf q$ component of the density in the relevant band, $U_c$ is the interaction strength, and $\Omega$ is the system area).
At a given density $n_{\rm tot}$, we compute the free energy as a function of the relevant order parameter $\hat{\phi}$, 
\begin{equation}
    F\left(\phi\right) = \left\langle H\right\rangle_{{\rm HF},\left(\phi,n_{\rm tot}\right)}
    -
    \left\langle H\right\rangle_{{\rm HF},\left(0,n_{\rm tot}\right)},\label{eq:FofPhiHF}
\end{equation}
where $\left\langle\right\rangle_{{\rm HF},\left(\phi,n\right)}$ is the Hartree-Fock expectation value for $\hat{\phi}=\phi$ at fixed density $n$.
We use the zero-temperature expression, as the temperature is assumed to be far below the $\hat{\phi}$ phase transition.
Contributions to Eq.~\eqref{eq:FofPhiHF} due to fluctuations around the mean-field solution, as well as due to other competing instabilities are beyond the current scope of this work.

As the energy scales associated with the order parameter jumps are comparable to the Fermi energy (see below) we will approximate the correlation length as the inter-particle separation, i.e., $\xi_{\phi}\approx n_{\rm tot}^{-1/2}$.
The stiffness is thus approximated by $\sigma\approx 8 \xi_\phi^2 F^*$, with the barrier height $F^*=\max F\left( \phi\in \left[0,\phi_0 \right]\right)$.
$\phi_0$ is the global minima of $F$, and we approximate $B \approx -F\left(\phi_0\right)$.

In the normal state, we evaluate two crucial quantities regarding superconductivity.
Namely, the critical temperature $T_c$, and the superconducting condensation energy, playing the role of $a^2/\left(4b\right)$ in our discussion above, Eq.~\eqref{eq:GLasymm}.
The superconducting coherence length is taken as a phenomenological parameters from experiments, and assumed to scale with $T_c$ in the conventional manner.
The details of our superconducting calculations are presented in Supplementary Section S2, where we use general considerations and avoid making assumptions on the origin of the pairing glue.
Moreover, we refrain from pinpointing the exact mechanisms by which the transition into the correlated phase extinguishes superconductivity, and keep our discussion as general as possible.
(In Supplementary Section S3 we discuss a curious scenario where the transition suppresses the superconducting phase through a combination of substantial DOS reconstruction and retardation effects.)

\textbf{Intervalley coherence in rhombohedral trilayer graphene (RTG).}
A promising candidate to observe the phenomenon introduced here is the superconductor region denoted as SC1 in Ref.~\cite{RTGsuperconductivityZhou2021}.
Its boundary in the $n_{\rm tot}$--$D$ plane (where $D$ is the perpendicular displacement field) coincides with a transition to a symmetry-broken correlated phase, which appears to be first-order~\cite{RTGsuperconductivityZhou2021,young_nadj_BBG_RTG_SC}.
Though the nature of the correlated phase terminating SC1 has not been confirmed, it is somewhat constrained.
The lack of spin and orbital ferromagnetism suggests either spin-valley locking, an intervalley-coherent phase (IVC), or combination thereof.
Theoretical studies have shown the IVC to be robust and ubiquitous throughout the phase diagram~\cite{RTG_Ashvin,RTG_Caltech}, steering our focus to the IVC case for simplicity.

\begin{figure}
    \centering
    \includegraphics[width=8.5cm]{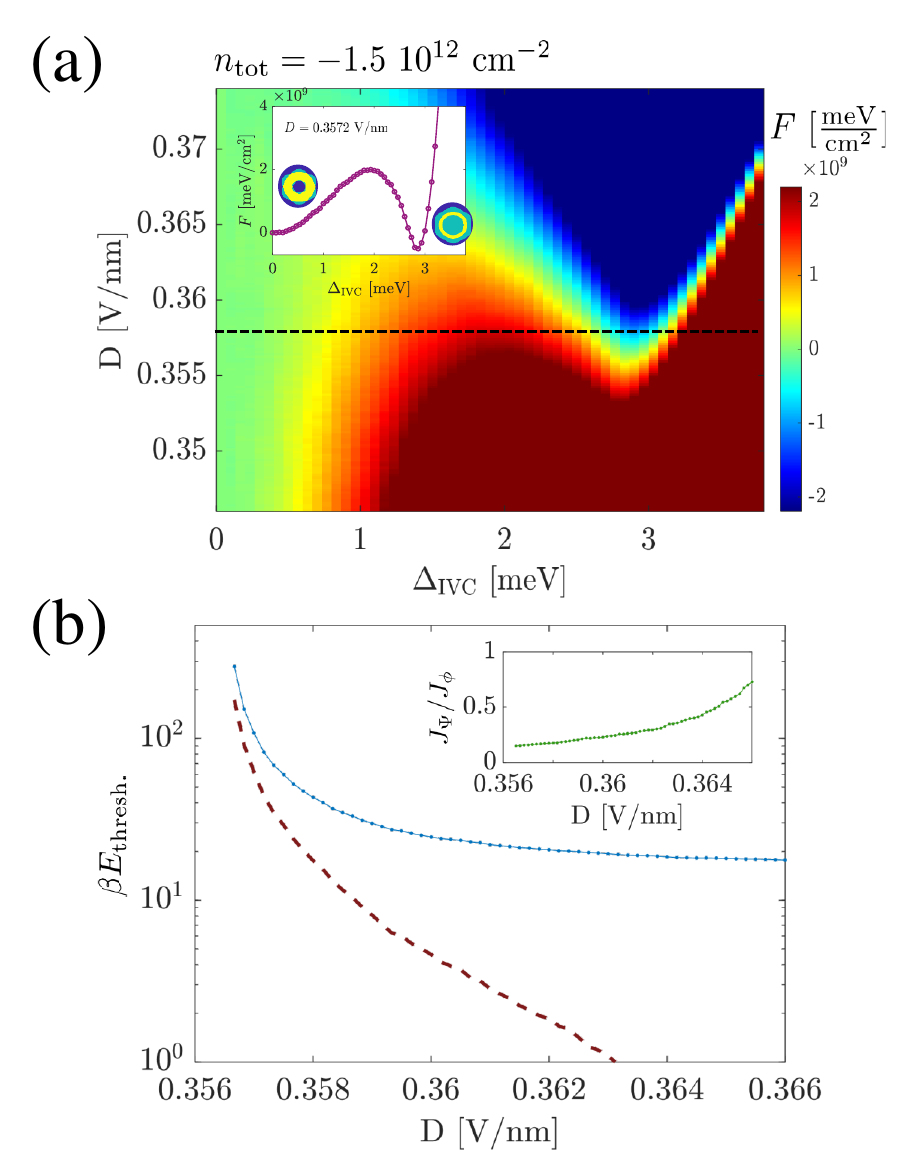}
    \caption{
    \textbf{Ephemeral superconductivity near an IVC transition in RTG.}
    (a)
    Free energy landscape for the IVC phase transition in RTG as a function of displacement field.
    A metastable regime is visible, where a free energy barrier separates an ordered minimum from the normal state.
    The fixed density is indicated.
    Inset: horizontal cut corresponding to the dashed black line in the main panel.
    The left and right circles are schematics of the Fermi surface in the normal and IVC phases, respectively.
    (b)
    The decay threshold energy in the vicinity of the transition (blue).
    The dashed purple line corresponds to the result in the absence of the superconductor.
    Inset: comparison of the surface tension contributions from the superconducting and IVC phases.
    The parameters used: $T=30$ mK $U_C=1.46$ eV nm$^2$, $g_{\rm att.}=0.21$ eV nm$^2$, $\omega^*=0.5$ meV.}
    \label{fig:IVCmetastability}
\end{figure}

In Fig.~\ref{fig:IVCmetastability}a we plot the Hartree-Fock energy landscape as a function of the IVC order parameter and displacement field.
Moving from low to high fields, one clearly observes a region of metastability: the global minimum is at $\sim 3$ meV, whilst the normal state remains locally stable.
At high enough values of $D$, the normal state finally becomes unstable.
We note in passing that in our Hartree-Fock calculations a similar region of metastability occurs when the displacement field as kept constant and the total density is varied, see Supplementary Section S3.

Next, we compute the false-vacuum decay threshold, shown in Fig.~\ref{fig:IVCmetastability}b.
As expected, close to the transition point it diverges, due to a vanishing energy difference between the FV and the true vacuum, driving the critical droplet size increasingly larger [Eq.~\eqref{eq:Rc}].
Notably, the threshold for decay becomes much smaller when the FV state is not superconducting, due to the surface tension contribution of the superconductor.
As shown in the inset, the relative strength of $J_\Psi$ increases, signaling two effects.
The first is a decrease in the ordered phase surface tension $J_\phi$ [Eq.~\eqref{eq:surfacetensionInstanton}] as the bias between the true and false vacuum increases.
The second is a result of \textit{enhanced superconducting $T_c$}, owing to an enhancement of DOS near the Fermi level as one moves deeper into the ordered state.

We now turn to estimate the life-time of this false vacuum,
\begin{equation}
    \tau_{\rm decay}\sim \tau_0 e^{\beta E_{\rm thresh.}},\label{eq:tau_decay}
\end{equation}
where $\tau_0$ is the much debated~\cite{Coleman_Quantum_PhysRevD.16.1762,MccumberHalperin} fluctuation time-scale prefactor.
For simplicity, we take the worst-case-scenario, and estimate it as the time two electrons separated by a correlation length can ``know about each other'',
$\tau_0\approx \xi_\phi / v_F \approx 10^{-13}$ sec.
Taking a reasonable $\beta E_{\rm thresh.}\sim 15-20$ leads to time scales
$ \tau_{\rm decay}\sim O\left(100 \,{\rm nsec} - 10\, \mu{\rm sec}\right)$.
For practical reasons, it is clear that $\tau_{\rm decay}$ should far exceed the timescale over which the quench in the external parameter is realized.
For the displacement field, the relevant timescale is the so-called RC charging time $\tau_{\rm RC}$ associated with the dual-gate geometry.
Using realistic device parameters, we estimate $\tau_{\rm RC}<0.1 {\rm nsec}$, orders of magnitude shorter than $\tau_{\rm decay}$.
We note that the estimated decay life-time \eqref{eq:tau_decay} assumes a single-droplet picture, neglecting intriguing possibilities of multi-droplet false vacuum decay and true vacuum percolation. This assumption is justified in the cases discussed here, where characteristic device sizes on the $\sim\mu$m scale, and are comparable to the typical size of the critical droplet $R_c$ in the vicinity of the transition (see Supplementary Section S3).

\textbf{Stoner blockade in Bernal bilayer graphene (BBG).}
The phase diagram of BBG has been shown to hosts a multitude of sharp phase transitions, as well as superconductivity in the presence of either an in-plane magnetic field or proximity to WSe$_2$~\cite{ZhouYoungBLGZeeman,nadj_ISOC_BBGBLGZhang2023,young_nadj_BBG_RTG_SC}.
The phenomenology is suggestive of a vicinity of a superconductive phase in the absence of these two perturbations, where the formation of a competing correlated phase suppresses it~\cite{Zhiyu_BBGPhysRevB.107.174512,StonerBlockadeBBGPhysRevB.108.024510}.
We consider an alternative route to circumvent the presumed correlated phase, by quenching across the transition.

\begin{figure}
    \centering
    \includegraphics[width=8.5cm]{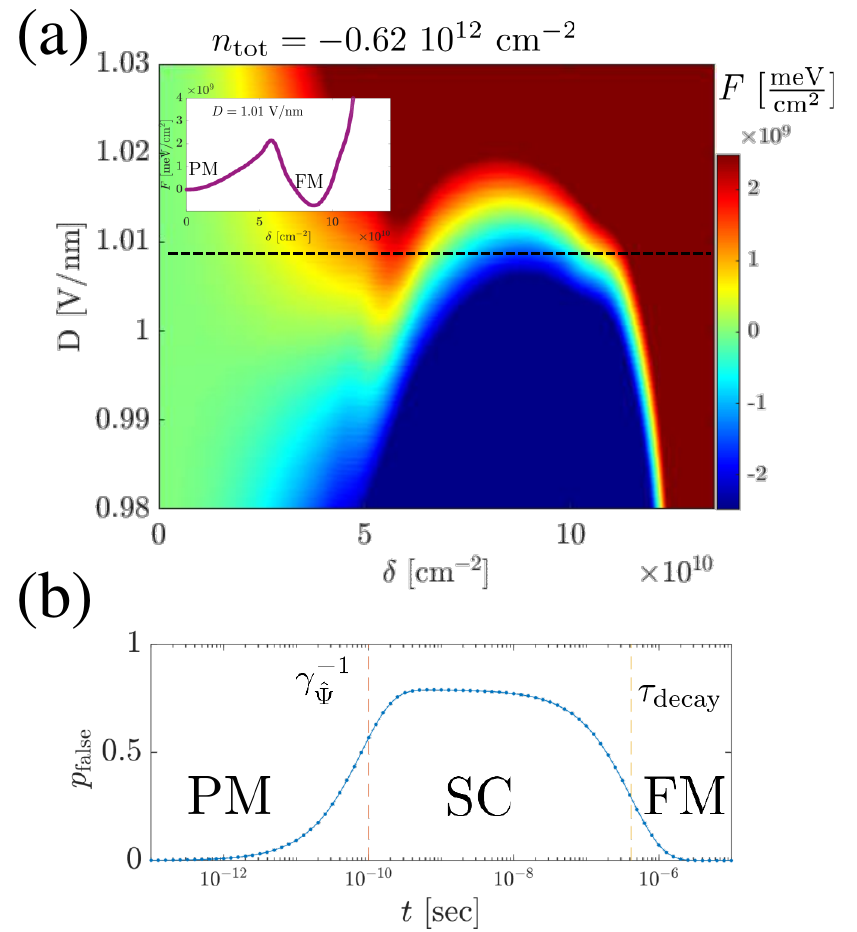}
    \caption{
    \textbf{Ephemeral superconductivity near a Stoner transition in BBG.}
    (a)
    Free energy landscape for the Stoner phase transition in BBG as a function of displacement field (the fixed density is indicated).
    Approaching the transition from the high $D$ side, a metastable regime is reached.
    Inset: horizontal cut corresponding to the dashed black line in the main panel.
    (b)
    Occupation probability of the false vacuum superconductor evolving in time [Eq.~\eqref{eq:heuristicmasterMain}].
    The system is initially in the FV paramagnetic state (PM), superconductivity develops atop the metastable FV (SC), followed by an eventual decay to the equilibrium ferromagnet (FM).
    The characteristic superconductor formation time $\gamma_{\hat{\Psi}}^{-1}$ and decay time are indicated by horizontal red and yellow lines, respectively.
    The parameters used: $T=30$ mK, $U_C=1.8$ eV nm$^2$, $g_{\rm att.}=0.265$ eV nm$^2$, $\omega^*=0.5$ meV,
    $t_{\rm decay}/\tau_0 = e^{15}$, $\gamma_0^{-1}/\tau_0 = e^{8}$, $\gamma_{\hat{\Psi}}^{-1}=100$ psec.
    }
    \label{fig:BBGmetastability}
\end{figure}

We focus on the vicinity of the magnetic-field-induced superconducting regime, where phenomenology is consistent with a fourfold to twofold degenerate Stoner transition.
The Hartree-Fock free energy as a function of the polarization order parameter is shown in Fig.~\ref{fig:BBGmetastability}a.
A metastable regime is clearly developed near the transition.
The normal-state energetic are quite similar to those obtained for the RTG case, and lead to similar energy threshold, even when accounting for a lower $T_c$ of the superconductor observed in experiments (see Supplementary Section S3).

There is a notable difference in this case compared to the RTG scenario above.
Here, the system is in the normal state after the quench.
It decays to the correlated phase with the characteristic $\tau_{\rm decay}$, yet superconductivity may form more rapidly, as it does not need to overcome an energy barrier.
We estimate the superconductor formation rate as $\gamma_{\hat{\Psi}}\sim \Delta_0$, i.e., roughly proportional to the equilibrium superconducting gap~\cite{Levitov_BCS_rate,Pekker_Demler_Rate}.
Combining the decay rate $\gamma=\tau_{\rm decay}^{-1}$, $\gamma_{\hat{\Psi}}$, and the decay rate in the absence of superconductivity $\gamma_0$, we may approximate the probability of the system being in the superconducting false vacuum (Supplementary Section S3),
\begin{equation}
    p_{{\rm false}}\left(t\right)=\frac{\gamma_{\hat{\Psi}}}{\gamma_{\hat{\Psi}}+\gamma_{0}-\gamma}\left[1-e^{-\left(\gamma_{\hat{\Psi}}+\gamma_{0}-\gamma\right)t}\right]e^{-\gamma t}.\label{eq:heuristicmasterMain}
\end{equation}
As demonstrated in Fig.~\ref{fig:BBGmetastability}b, at early times superconductivity builds up at a rate $\sim\gamma_{\hat{\Psi}}$ to some finite fraction, and decays at long times at the rate $\gamma$.
Our calculations indicate superconductivity should remain visible up to $O\left(100\right)$ nsec timescales.

In principle, for the specific scenario described in this section, where the system is quenched rapidly through a superconducting phase transition, one should take into account the formation of topological defects (vortices) due to the Kibble-Zurek mechanism~\cite{KIBBLE1980183,Zurek1985}.
Physically, one expect that following the quench, independently coherent domains of size $\sim \xi_{\Psi}$ form,
and coalesce at a time-scale
$\tau_{GL}\approx\pi/\left(8\left|T_c-T\right|\right)^{-1}$~\cite{non_eq_SC_Kopnin,BEC_vortices_Weiler2008}, the Ginzburg-Landau relaxation time~\cite{larkin2005theory}.
We note that $\tau_{GL}$ is of the same order of $\gamma_{\hat{\Psi}}^{-1}$.
The relaxation dynamics of these defects, as well as oscillations in the order parameter magnitude~\cite{Levitov_BCS_rate} are presumed to play a secondary role, and are thus beyond the current scope of this work.

\subsection*{Detecting ephemeral superconductivity
} \label{sec:detectability}

Thus far we have demonstrated the stabilization of the false vacuum by superconductivity, as well as the potential enhancement of the superconducting state in the metastable regime.
We now explore another intriguing consequence of the physics studied in this work.
Namely, the superconducting nature of the false vacuum make the phenomenon readily accessible to transport measurements.

\begin{figure*}
    \centering
    \includegraphics[width=16cm]{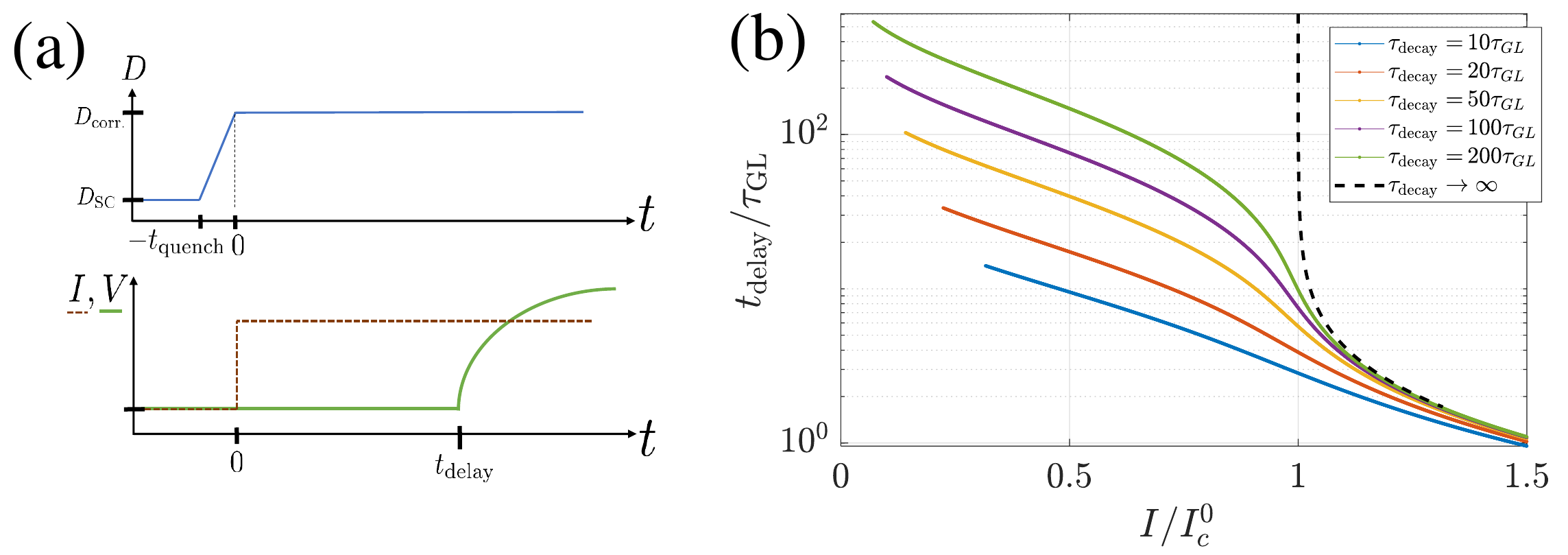}
    \caption{
    \textbf{Measuring the order parameter relaxation dynamics.}
    (a)
    A control parameter, e.g., $D$ is quenched through the first order phase transition (top).
    The quench duration is $t_{\rm quench}$, and it ends at $t=0$, at which time a current is driven through the system and the time-resolved voltage signal is measured (bottom).
    (b) Calculations of the order parameter relaxation, performed by integrating Eq.~\eqref{eq:relaxationFULLMain} over time.
    Each curve is terminated at small currents at a value
    $I= I_{c}^0\sqrt{\frac{\tau_{GL}}{\tau_{\rm decay}}}$, where we expect our simplified treatment to no longer be a good approximation (Supplementary Section S4).
    }
    \label{fig:detectabilitySC}
\end{figure*}

Consider an experimental setup similar to the one pioneered in Ref.~\cite{relaxation_SC_PALS1979150}, depicted in the inset of Fig.~\ref{fig:detectabilitySC}.
The tuning parameter in the system, e.g., displacement field, is quenched from time $-t_{\rm quench}$ to $t=0$.
At $t=0$ a current pulse is fed to the device, and the transient voltage response is recorded.
At a time $t=t_{\rm delay}$ voltage drop is eventually observed.
originally, the experiment was tailored to measure the relaxation time of equilibrium superconductors, and thus no signal would appear if the amplitude of the current pulse is less than the superconducting critical current $I_c$.
Here, the situation is made significantly richer by the decay of superconductivity itself.

Under some simple yet reasonable assumptions (see Supplementary Section S4), we find the delay as the time at which $f\left(t_{\rm delay}\right)=0$, with the initial conditions $f\left(0\right)=1$, where $f$ obeys the time-evolution,  
\begin{equation}
    \tau_{GL}\partial_{t}f=-e^{2t/\tau_{{\rm decay}}}\left(\frac{I}{I_{c}}\right)^{2}\frac{4}{27f^{3}}+f\left(1-f^{2}\right).\label{eq:relaxationFULLMain}
\end{equation}

In Fig.~\ref{fig:detectabilitySC} we demonstrate the dependence of the voltage signal delay time on the amplitude of the injected current.
If the false vacuum state is stable enough, a noticeable uptick in the delay is made clearly visible for currents slightly below the equilibrium critical values.
At relatively small values of the probe current, the expected delay is of the same order as the false-vacuum decay time $\tau_{\rm decay}$.

\section*{Discussion}\label{sec:conclusions}

Ephemeral superconductivity is a fascinating non-equilibrium state of matter, which is yet to be fully clarified~\cite{non_eq_SCgalitskii1973feasibility,non_eq_SC_ExcessQP_Elesin_1981,non_eq_SC_Kopnin,spinglassVS_SC,non_eq_SC_Chamon,SCenhancementbymodulation,Cavalleri_Cuprate_10psec}.
Here, we explore an unusual striking phenomenon: Superconductivity developing on top of a correlated false-vacuum manifold.
In contrast with previous scenarios where superconductivity is terminated by a first-order phase transition~\cite{PhysRevLett.firstorderSC_cecoin5,firstorderSC_sr2ru04}, the transient superconductor is temporarily protected by the correlated FV.
Furthermore, the FV can facilitate the formation of a superconductor followed bu its decay (Fig.~\ref{fig:BBGmetastability}b), in otherwise superconductivity-excluded areas of the equilibrium phase diagram.
The proposed non-equilibrium superconductor is incompatible with the true many-body ground state of the system, and is thus ephemeral albeit metastable.

We elucidated how this FV superconductor can remarkably and significantly \textit{enhance the false-vacuum lifespan}, possibly by a few orders of magnitude.
This enhancement occurs even if the superconducting condensation energy is negligible as compared to that of the correlated true vacuum, as we demonstrated by utilizing a general phenomenological framework [Eq.~\eqref{eq:GLasymm}].
This unique effect is rooted in a notable surface tension contribution of the superconductor to the nucleation of critical threshold droplets, which facilitate the decay~\cite{LANGER1967108,Coleman_SemiclassicalPhysRevD.15.2929,altland2010condensed}.
The prospect of FV stabilization by a generic secondary order, which is adversarial to the corresponding true vacuum, is left for future work, as well as a broader view on quantum effects and the hybrid quantum-classical regime.

We demonstrated how the ephemeral superconducting state can be realized in multilayer graphene devices.
These systems ubiquitously show the coexistence of superconducting regions in the phase diagram, and apparent first-order phase transitions to correlated symmetry-broken phases~\cite{RTG_half_quarter_metal_Zhou2021,young_nadj_BBG_RTG_SC,Nadj_zhang2024twistprogrammablesuperconductivityspinorbitcoupled,LongJuyang2024diverseimpactsspinorbitcoupling,Young_patterson2024superconductivityspincantingspinorbit}.
On occasions, these transitions actually terminate the superconducting dome, making such devices an ideal platform to explore the intertwined false-vacuum physics.
Combining Hartree-Fock calculations for the symmetry-broken transition energetics, theory of the superconductivity transition, and phenomenology obtained from experiments, we estimate the relevant FV  lifetime for observing the FV decay to be $\gtrsim O\left(100\,{\rm nsec}\right)$,
rendering the decay process as much slower as compared to recently observed transient superconducting phenomena~\cite{Cavalleri_Cuprate_10psec,Cavalerri_PhysRevB.91.020505,Cavalleri_K3C60,photo_SCIsoyama2021,Cavalerri_resonantRowe2023}. 

Another intriguing prospect in these graphene systems, is that competition with a correlated phase transition arguably undermines the potential equilibrium superconducting orders.
(This is reminiscent of the competing phases in high-$T_C$ superconductors, e.g.,~\cite{cuprate_cdw_sc_competition_Chang2012}.)
The multilayer graphene superconductors were observed to be stabilized by some perturbations, e.g., magnetic fields or proximity to a non-trivial substrate.
However, we have pointed out a different route for bringing these low-lying superconductors to light: transiently suppressing the symmetry-broken state by quenching the system through the first-order transition.
The motivation for such a process to be feasible is simple.
The same factor that participates in driving the transitions, the DOS near the Fermi energy, is the one that is expected to enhance superconductivity when given the chance (see related discussion in Refs.~\cite{Shavit_MatbgPhysRevLett.127.247703,StonerBlockadeBBGPhysRevB.108.024510}).

An important consequence of ephemeral superconductivity ``dressing'' the metastable false-vacuum is that direct experimental detection of its decay becomes simple and experimentally accessible.
The long time-scales for the superconducting life-times mentioned above are in fact conducive to straightforward transport measurements.
Resolution of $\sim 1$ nsec in the voltage-delay experimental scheme we detail in Sec.~\ref{sec:detectability} can readily be obtained.
Tracking the temporal signal due to the decay of the superconductor, and as a consequence the establishment of the equilibrium ground-state, may be achieved in these graphene systems (as well as in another candidate system, twisted WSe$_2$~\cite{Wse2_hysteresis_CoryDean_guo2024superconductivity}), without the need of sophisticated light-based apparatus.
Moreover, conclusive transport signatures of the kind we discussed obviate many of the ambiguities associated with photo-measurements of non-equilibrium superconductors~\cite{PatrickLee_SC_like_PhysRevB.104.054512}. 
These facts, along with the ability to electrically tune them across the phase diagram, make highly-correlated multilayer graphene devices ideal playgrounds for manipulating and probing the false-vacuum decay in the context of solid-state systems.

\section*{Data Availability}
The data supporting the findings in this work are publicly available online at \url{https://doi.org/10.5281/zenodo.14290558}.

\section*{Code Availability}
The code used to generate the data and figures appearing in this work is available online at \url{https://github.com/galshavit/EphemeralSC}.

\bibliographystyle{aapmrev4-2}
\bibliography{metaSC}

\begin{acknowledgments}
We thank John Curtis, Thomas Weitz, Dmitry Abanin, and Erez Berg for enlightening discussions.
S.N.-P. acknowledges support from the National Science Foundation (grant number DMR-1753306). G.R. and S.N.-P. also acknowledge the support of the Institute for 
Quantum Information and Matter, an NSF Physics Frontiers 
Center (PHY-2317110).
G.S. acknowledges support from the Walter Burke Institute for Theoretical Physics at Caltech, and from the Yad Hanadiv Foundation through the Rothschild fellowship.
Part of this work was done at the Aspen Center for Physics, which is supported by the NSF grant PHY-1607611.
\end{acknowledgments}

\section*{Author Contributions}
G.S. and G.R. conceived the idea for this project.
G.S., S.N.-P., and G.R. performed research, analyzed data and calculation results, and wrote the manuscript.

\section*{Competing Interests}
The authors declare no competing interests.

\begin{widetext}
\section*{Supplementary Information for "Ephemeral Superconductivity Atop the False Vacuum"}

\global\long\def\thesection{S\Alph{section}}%
 \setcounter{figure}{0} 
\global\long\def\thefigure{S\arabic{figure}}%
 \setcounter{equation}{0} 
\global\long\def\theequation{S\arabic{equation}}%

\setcounter{section}{0} \renewcommand{\thesection}{S\arabic{section}} 
\setcounter{figure}{0} \renewcommand{\thefigure}{S\arabic{figure}} 
\setcounter{equation}{0} \renewcommand{\theequation}{S\arabic{equation}}

\setcounter{figure}{0}
\setcounter{section}{0}
\renewcommand{\figurename}{Supplementary Fig.}

\section{Phenomenological model calculations}
We discuss and analyze the following Ginzburg-Landau free energy functional from the main text, 
\begin{align}
    F	&=\int d^{2}x\left[\frac{\sigma}{2}\left|\nabla\phi\right|^{2}+16g\phi^2\left(\phi-1\right)^{2}-B\phi^2\right]\nonumber\\
	&+\int d^{2}x\left[\frac{\kappa}{2}\left|\nabla\Psi\right|^{2}-\frac{a\left(r\right)}{2}\left|\Psi\right|^{2}+\frac{b}{4}\left|\Psi\right|^{4}\right] \nonumber\\
	&+\frac{\lambda}{2}\int d^{2}x\left|\Psi\right|^{2}\phi^{2}. \label{eq:GLasymm}
\end{align}
\subsection*{Surface tension calculations}\label{appsec:phenosurface tension}
Given the free energy
\begin{equation}
    F_\phi=\int d^{2}x\left[\frac{\sigma}{2}\left|\nabla\phi\right|^{2}+16g\phi^{2}\left(\phi-1\right)^{2}\right],
\end{equation}
we calculate the energy associated with a domain-wall solution of extent $\xi$, $\phi\left(x\right)=\frac{1}{2}\left({\rm tanh}\frac{x}{\xi}+1\right)$ per unit length along the domain wall direction $L_y$,
\begin{equation}
    \frac{F_\phi}{L_{y}}=\left[\frac{\sigma}{8\xi}+g\xi\right]\int_{-\infty}^{\infty}dz\frac{1}{\cosh^{4}z}.
\end{equation}
Minimizing with respect to $\xi$, we obtain the correlation length
$\xi_\phi=\sqrt{\frac{\sigma}{8g}}$, and the surface tension
$J_\phi^0=\frac{8}{3}g\xi_\phi=\sqrt{\frac{8\sigma g}{9}}$.

When an additional bias is added to $F_\phi$ favoring the ordered phase this surface tension is only approximately correct.
The bias always tends to lower the surface tension, and can generally be calculated by integrating over the effective equations of motion, i.e.,
\begin{equation}
    J_\phi = \int_0^{\phi_{>0}} d\phi\sqrt{2\sigma V\left(\phi\right)}
    = 4g\xi_\phi\int_0^{\phi_{>0}} d\phi\sqrt{V\left(\phi\right)/g},\label{eq:appsurfacetensionInstanton}
\end{equation}
where $V\left(\phi\right)=16g\phi^{2}\left(\phi-1\right)^{2}-B\phi^2$, and $\phi_{>0}$ is the maximal $\phi$ for which $V\left(\phi\right)>0$.
In the right hand side of Supplementary Equation~\eqref{eq:appsurfacetensionInstanton} we have expressed the surface tension in terms of the parameters $g$ and $\xi_\phi$, which we can later input from either phenomenology or mean-field calculations.
One easily verifies that for $B=0$ one recovers exactly $J_\phi^0$.

Repeating the first part of the calculation above, now for the $\hat{\Psi}$ part, with the free energy
\begin{equation}
    F_\Psi=\int d^{2}x\left[\frac{\kappa}{2}\left|\nabla\Psi\right|^{2}-\frac{a}{2}\left|\Psi\right|^{2}+\frac{b}{4}\left|\Psi\right|^{4}\right],
\end{equation}
we obtain the correlation length
$\xi_\Psi = \sqrt{\frac{2\kappa}{a}}$, and the surface tension
$J_\Psi = \frac{8}{3}\xi_\Psi\frac{a^{2}}{4b}$.

\subsection*{Explicit results from the phenomenological model}

In Supplementary Fig.~\ref{fig:RcBarrierPheno_extra} we demonstrate how the critical bubble radius and the value of the energy threshold depend on the metastability bias parameter $B$.
Clearly, as the ordered $\phi\approx 1$ phase becomes increasingly favorable, true vacuum bubbles become easier to nucleate.
The impact of the competing $\hat{\Psi}$ phase, and the enhancement of metastability due to its presence are quite clear -- both $R_c$ and $E_{\rm thresh.}$ increase substantially as compared to the case of its absence. 

\begin{figure}
    \centering
    \includegraphics[width=17cm]{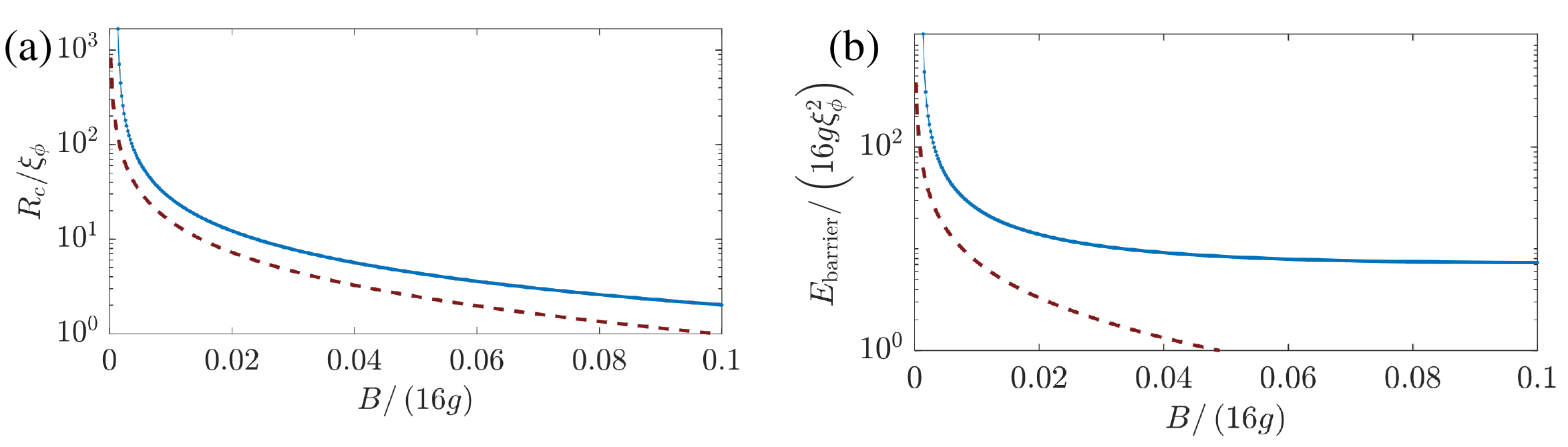}
    \caption{ 
    (a)--(b) calculated $R_c$ and $E_{\rm thresh.}$ (blue lines) using the phenomenological theory, Supplementary Equation~\eqref{eq:GLasymm}.
    The dashed purple lines denote the respective results in the absence of the $\hat{\Psi}$ phase.
    Here we parameterized $\xi_\psi=20\xi_\phi$, $a=a_0+\alpha B$, $b=a_0=16g/200$, and $\alpha=0.006$.
    }
    \label{fig:RcBarrierPheno_extra}
\end{figure}

\subsection*{Quantum regime}\label{sec:appquantumcorrestions}
It is instructive to explore the false vacuum decay of the $\hat{\phi}$ phase in arbitrary dimensionality $d$, and with possibly important quantum fluctuations. 
For that end, we introduce the imaginary time action
\begin{equation}
    S_{\phi}=\int d\tau\int d^{d}x\left[4g\tau_{\phi}^{2}\left(\partial_{\tau}\phi\right)^{2}+4g\xi_{\phi}^{2}\left|\nabla\phi\right|^{2}+16g\phi^{2}\left(\phi-1\right)^{2}-B\phi^{2}\right],
\end{equation}
where we have re-written the $\sigma$ term from Supplementary Equation~\eqref{eq:GLasymm} in terms of the correlation length, and $\tau_{\phi}$ can be thought of as a correlation time characterizing quantum fluctuations. 
For example, in the limit $\tau_{\phi}\to\infty$, fluctuations with respect to $\tau$ freeze out, and we may ignore the $\left(\partial_{\tau}\phi\right)^{2}$ contribution entirely.
Thus one replaces the integration over the Lagrangian $\int d\tau{\cal L}_\phi$, by a simple factor of the ``length'' of the system in imaginary-time, $\beta{\cal L}_\phi$, i.e., we recover the classical expression.

To simplify the analysis, we rescale the action $S_{\phi}$ as the following, $x\to x\xi_{\phi}$, $\tau\to\tau\tau_{\phi}$,
\begin{equation}
    \frac{S_{\phi}}{\tau_{\phi}\xi_{\phi}^{d}}=\int d^{d+1}x\left[4g\left|\nabla_{d+1}\phi\right|^{2}+16g\phi^{2}\left(\phi-1\right)^{2}-B\phi^{2}\right].
\end{equation}
With this isotropic form, one may readily employ the bubble formalism we use in the main text. 
The action barrier of forming a true vacuum bubble is approximately given by
\begin{equation}
    \frac{S_{{\rm bubble}}}{\tau_{\phi}\xi_{\phi}^{d}}=\frac{\pi^{\frac{d+1}{2}}}{\Gamma\left(\frac{d+1}{2}+1\right)}\left[-R^{d+1}B+\left(d+1\right)R^{d}\tilde{J_{\phi}}\right],
\end{equation}
where $\Gamma$ is Euler's gamma function, and the rescaled surface tension is $\tilde{J_{\phi}}=16g\int_{0}^{\phi_{>0}}d\phi\sqrt{\phi^{2}\left(\phi-1\right)^{2}-\frac{B}{16g}\phi^{2}}$.
The critical (dimensionless) $\tilde{R_{c}}$ is then found to simply be 
\begin{equation}
    \tilde{R_{c}}=\frac{d\tilde{J_{\phi}}}{B}.
\end{equation}
We thus find the action barrier for the formation of the critical bubble,
\begin{equation}
    S_{{\rm Q}}=\frac{\pi^{\frac{d+1}{2}}d^{d}}{\Gamma\left(\frac{d+1}{2}+1\right)}\tau_{\phi}\left(\frac{\tilde{J_{\phi}}}{B}\right)^{d}\xi_{\phi}^{d}\tilde{J_{\phi}}.
\end{equation}
Going through the same steps for the classical case ($\tau_{\phi}\to\infty$), we find the classical barrier under the same conditions,
\begin{equation}
    S_{{\rm C}}=\frac{\pi^{\frac{d}{2}}\left(d-1\right)^{d-1}}{\Gamma\left(\frac{d}{2}+1\right)}\beta\left(\frac{\tilde{J_{\phi}}}{B}\right)^{d-1}\xi_{\phi}^{d}\tilde{J_{\phi}}.
\end{equation}
One clearly sees that due to the different dependence on $\tilde{J_{\phi}}/B$, the classical energy threshold is much lower close to the transition ($B\ll\tilde{J_{\phi}}$), and thus thermal fluctuations dominate. 
At low enough temperatures however, the $\beta$ prefactor can become significant, and quantum decay of the false vacuum ``short-circuits'' the classical path. 
We may estimate the temperature below which this becomes important,
\begin{equation}
    T_{Q}\approx\frac{\Gamma\left(\frac{d+1}{2}+1\right)\left(d-1\right)^{d-1}}{\sqrt{\pi}\Gamma\left(\frac{d}{2}+1\right)d^{d}}\frac{B}{\tilde{J_{\phi}}}\tau_{\phi}^{-1}\to_{d=2}\frac{3}{32}\frac{B}{\tilde{J_{\phi}}}\tau_{\phi}^{-1},
\end{equation}
and on the right hand side we explicitly plugged in $d=2$, which is appropriate for our considerations in this work. 
The upshot here is that close enough to the transition, one may find a regime where the classical decay paths are most important.

Let us make another comment on scenarios where $\tau_{\phi}$ may be rather pathological. 
In cases where the order parameter commutes with the Hamiltonian, on a mean-field level one expects to find $\tau_{\phi}^{-1}=0$, and the false vacuum can only decay through thermal fluctuations.
This is the case for example for conventional Stoner transitions, and for the transverse field Ising model with vanishing transverse field. 
Realistically, due to corrections and phenomena beyond our simplified descriptions, $\tau_{\phi}^{-1}$ may be finite, yet one expects it to be rather suppressed as compared to other generic symmetry breaking transitions, e.g., an IVC order.

\subsection*{Hybrid regime}
In the presence of both the $\hat{\phi}$ and $\hat{\Psi}$ phases in the false vacuum decay problem, there exists another regime of interest.
Assume that the temperature is below $T_Q$, which is set by the properties of the $\hat{\phi}$ action, yet still significantly higher than the temperature where variations of $\Psi$ in imaginary time $\tau$ are enabled.
Namely,
\begin{equation}
    T_Q\frac{\tau_\phi}{\tau_\Psi}<T<T_Q.
\end{equation}

Let us simplify the analysis in this regime by limiting ourselves to the $d=2$ case.
The critical bubble will assume an anomalous shape here.
Its core would be comprised of an ellipsoid with radius $\sim \tilde{R_c}\xi_\phi$ and height $\sim2\tilde{R_c}\tau_\phi$ (we have simplified even more, by assuming the $\hat{\Psi}$ action has little effect on the value of the critical $\tilde{R_c}$).
However, this ellipsoid would be embedded inside a cylinder of height $\beta$ and radius $\sim\tilde{R_c}\xi_\phi+\delta R$ of the intermediate $\phi=0$, $\Psi=0$ state, akin to the corona in Supplementary Fig.~\ref{fig:bubblefigure}a.
Since $\tau_\Psi$ is too large, the $\hat{\Psi}$ sector still attempts to avoid variations in the imaginary time direction, leading to the cylindrical shape of the domain wall.
Under these assumptions, the action barrier $S_Q^\Psi$ now reads
\begin{equation}
    S_{Q}^{\Psi}=S_{Q}\left[1+8\frac{T_{Q}}{T}\left(1+\frac{\xi_{\Psi}}{\xi_{\phi}}\frac{B}{4\tilde{J_{\phi}}}\right)^{2}\frac{a^{2}}{4b}/B\right].
\end{equation}
Notice that even if the ratio $\frac{a^2}{4b}/B$ is rather small (as in the scenarios we considered in the main text), a significant stabilization of the false vacuum decay may occur due to the combination of the large prefactors $8T_Q/T$ and $\xi_\Psi/\xi_\phi$.
For example, taking the reasonable parameters $B=0.4\tilde{J_\phi}$, $\xi_\Psi=20\xi_\phi$, and $T_Q=5T$ one finds 
$S_{Q}^{\Psi}=S_{Q}\left[1+360\frac{a^{2}}{4b}/B\right]$.
It thus becomes plausible in this regime that the $\hat{\Psi}$ contribution to the metastability becomes comparable and possibly dominates over the $\hat{\phi}$ contribution, even if the discrepancy in energy scales is very large.

\section{Microscopic calculations}

\subsection*{Crystalline graphene band structure}

We begin by calculating the the non-interacting band structure of biased Bernal-stacked bilayer graphene. 
Expanded around the valley $K/K'$ points in momentum space, the Hamiltonian is~\cite{McCann_2013BLG} 
\begin{equation}
H_{BLG}=\sum_{\mathbf{k},\tau,s}c_{\tau s\mathbf{k}}^{\dagger}h_{\tau}\left(\mathbf{k}\right)c_{\tau s\mathbf{k}},\label{eq:BLGhAMILTONIAN}
\end{equation}
with $c_{\tau s\mathbf{k}}=\left(A_{1,\tau s\mathbf{k}},B_{1,\tau s\mathbf{k}},A_{2,\tau s\mathbf{k}},B_{2,\tau s\mathbf{k}}\right)^{T},$
where $X_{i,\tau s\mathbf{k}}$ annihilates an electron on sub-lattice
$X$ in layer $i$, with spin $s$, and momentum $\mathbf{k}$ near
the valley $\tau$. The matrix $h_{\tau}$ is given by
\begin{equation}
h_{\tau}\left(\mathbf{k}\right)=\begin{pmatrix}\frac{U}{2} & v_{0}\pi_{\tau}^{*} & -v_{4}\pi_{\tau}^{*} & -v_{3}\pi_{\tau}\\
v_{0}\pi_{\tau} & \frac{U}{2}+\Delta' & \gamma_{1} & -v_{4}\pi_{\tau}^{*}\\
-v_{4}\pi_{\tau} & \gamma_{1} & -\frac{U}{2}+\Delta' & v_{0}\pi_{\tau}^{*}\\
-v_{3}\pi_{\tau}^{*} & -v_{4}\pi_{\tau} & v_{0}\pi_{\tau} & -\frac{U}{2}
\end{pmatrix},\label{eq:H04by4basis}
\end{equation}
with $\pi_{\tau}=\tau k_{x}+ik_{y}$, and the parameters $v_{i}=\frac{\sqrt{3}}{2}a\gamma_{i},$
$a=0.246$ nm, $\gamma_{0}=2.61$ eV, $\gamma_{1}=361$ meV, $\gamma_{3}=283$
meV, $\gamma_{4}=138$ meV, and $\Delta'=15$ meV \citep{tbParameters}.
The interlayer potential difference $U$ is approximately $U\approx-dD/\epsilon$,
where the interlayer distance is $d\approx0.33$nm, $\epsilon\approx4.3$,
and $D$ is the displacement field. 
We diagonalize $H_{BLG}$ at each momentum, and extract the dispersion relation of the lowest-lying valence band, which we denote by $\epsilon_{\tau,\mathbf{k}}$.

For the rhombohedral trilayer graphene calculations, we follow a similar method.
We diagonalize the Hamiltonian adapted from Refs.~\cite{ABCparametersTB_mccan,ABCparametersTB},
\begin{equation}
    h_\tau\left({\bf k}\right)=\begin{pmatrix}\Delta+\delta_{1}+\delta_{2} & \frac{\gamma_{2}}{2} & v_{0}\pi^{\dagger} & v_{4}\pi^{\dagger} & v_{3}\pi & 0\\
\frac{\gamma_{2}}{2} & -\Delta+\delta_{1}+\delta_{2} & 0 & v_{3}\pi^{\dagger} & v_{4}\pi & v_{0}\pi\\
v_{0}\pi & 0 & \Delta+\delta_{2} & \gamma_{1} & v_{4}\pi^{\dagger} & 0\\
v_{4}\pi & v_{3}\pi & \gamma_{1} & -2\delta_{2} & v_{0}\pi^{\dagger} & v_{4}\pi^{\dagger}\\
v_{3}\pi^{\dagger} & v_{4}\pi^{\dagger} & v_{4}\pi & v_{0}\pi & -2\delta_{2} & \gamma_{1}\\
0 & v_{0}\pi^{\dagger} & 0 & v_{4}\pi & \gamma_{1} & -\Delta+\delta_{2}
\end{pmatrix},
\end{equation}
where $\Delta$ is the interlayer potential difference (proportional to the displacement field $D$.
Notice we have written the Hamiltonian in the basis $\left(A_1,B_3,B_1,A_2,B_2,A_3\right)$, where $A_j/B_j$ correspond to different graphene sublattices in layer $j$.
Here, we use the adopt the parameters $\gamma_0=3.1$ eV, $\gamma_1=0.38$ eV, $\gamma_2=-15$ meV, $\gamma_3=-0.29$ eV, $\gamma_4=-141$ meV, $\delta_1=-10.5$ meV, $\delta_2=-2.3$ meV.
Frot the trilayer calculations in this work, we once again are only interested in the dispersion relation of the lowest lying valence band.

\subsection*{IVC phase transition}\label{sec:appIVC}
We consider the phase transition at a given total density n from a normal symmetric phase, where each of the four spin-valley flavors has occupation $n/4$, and between an intervalley coherent (IVC) phase. 
The non-interacting spectrum of the relevant valence band at the $\tau=\pm$ valley is $\epsilon_{\tau}\left({\bf k}\right)$, and depends on the value of the displacement field. 

For simplicity, we assume the IVC is doubly degenerate, and no further symmetry breaking. Denoting the momentum-independent IVC order parameter $\Delta_{{\rm IVC}}$, we can calculate the mean field free-energy with respect to the normal state,

\begin{equation}
    F_{{\rm IVC}}\left(\Delta_{{\rm IVC}}\right)=2\left[F_{{\rm kin}}\left(\Delta_{{\rm IVC}}\right)-F_{{\rm kin}}\left(\Delta_{{\rm IVC}}=0\right)+\Omega\frac{\Delta_{{\rm IVC}}^{2}}{g}\right],
\end{equation}
with $\Omega$ the system volume, $g$ the relevant coupling constant
in the IVC channel, the factor of $2$ accounts for the two-fold degeneracy,
and
\begin{equation}
    F_{{\rm kin}}\left(\Delta_{{\rm IVC}}\right)=\sum_{\mathbf{k},i}E_{\mathbf{k},i}\,\Theta\left[\mu\left(\Delta_{{\rm IVC}}\right)-E_{\mathbf{k},i}\right].
\end{equation}
Here, the mean field energies are 
\begin{equation}
    E_{\mathbf{k},1/2}=\frac{\epsilon_{+}+\epsilon_{-}}{2}\pm\sqrt{\left(\frac{\epsilon_{+}-\epsilon_{-}}{2}\right)^{2}+\Delta_{{\rm IVC}}^{2}},
\end{equation}
and the chemical potential $\mu\left(\Delta_{{\rm IVC}}\right)$ is
determined self-consistently by the relation 
$\frac{2}{\Omega}\sum_{\mathbf{k},i}\Theta\left[\mu\left(\Delta_{{\rm IVC}}\right)-E_{\mathbf{k},i}\right]=n$.

\subsection*{Stoner transition}
Here, we outline the calculation of the free energy in the vicinity of the first-order phase transitions we consider in this work.
We simplify, by assuming the relevant energetics do not change much in the sub-Kelvin temperature ranges (where superconductivity is measured in experiments).
This is justified if the transition temperature associated with these transitions is on the order of several Kelvin or higher, which appears to be experimentally consistent~\cite{ZhouYoungBLGZeeman,RTG_half_quarter_metal_Zhou2021}.

Assuming a short-range density-density interaction of strength $U$, the system can lower its energy on a mean-field level by re-distributing the electronic densities between different flavors with density $n_i$
\begin{equation}
    E_{\rm int} = \frac{U}{2}\left[\left(\sum_i n_i\right)^2 - \sum_i n_i^2\right].
\end{equation}
We will mostly be interested in transitions from a fully flavor-symmetric $2d$-fold degenerate state, to a $d$-fold degenerate state.
Concretely, in graphene systems with valley degeneracy, $d=2$, and the transition we focus on corresponds to, e.g., spin polarization or spin-valley locking.
Thus, fixing the total density $n_{\rm tot}$, one has $d$ flavors at density $n_+=\frac{n_{\rm tot}}{2d}+\delta$, and $d$ flavors with $n_-=\frac{n_{\rm tot}}{2d}-\delta$.
Thus, the contribution from the spontaneous polarization to the free energy density is
\begin{equation}
    F_{\rm int}/d = - U \delta^2.
\end{equation}

Whereas the interaction $U$ favors the Stoner transition, the kinetic energy associated with the band disfavors it.
This can be quantified by the energy cost associated with populating (depopulating) states with higher (lower) energies as compared to the non-interacting Fermi level,
\begin{equation}
    F_{\rm kin}/d=\left[\int_{E_{F}}^{E_{+}} - \int_{E_{-}}^{E_{F}}\right]
     d\epsilon{\cal N}\left(\epsilon\right)
     \left(\epsilon-E_F\right),
\end{equation}
where ${\cal N}$ is the non-interacting DOS, and the Fermi energies are obtained by the relations
$\int_{-\infty}^{E_F} d\epsilon{\cal N}\left(\epsilon\right) =\frac{n_{\rm tot}}{2d}$, 
$\int_{-\infty}^{E_\pm} d\epsilon{\cal N}\left(\epsilon\right) =\frac{n_{\rm tot}}{2d}\pm \delta$. 

\subsection*{Superconductivity calculations}\label{secapp:SC}
The nature of the superconducting order parameter, as well as the origin of the pairing glue, is not yet resolved in the graphene materials we discuss in the main text.
Further more, different materials and different superconducting regions in the phase diagram of the same device may have different properties and origins.
We thus refrain from an attempt to clarify such important questions regarding these superconductors, and keep our discussion as general as possible.

We assume some retardation energy scale, $\omega^*$, at which an attractive interaction in the Cooper channel $g_{\rm att.}$ is introduced, and restrict ourselves to a superconducting gap with trivial symmetry.
Accounting for the suppression of the initial repulsive interaction in the Cooper channel $U_0$ by the Anderson-Morel mechanism~\cite{AndersonMorel,tolmachev1962logarithmic}, we solve the following equation for $T_c$,
\begin{equation}
    \left(g_{\rm att.}-\frac{U_0}{1+U_0\ell}\right){\cal D}=1,\label{eq:findTcTAM}
\end{equation}
where
\begin{equation}
     \ell = \int_{\omega^*}^\infty d\xi \frac{{\cal N}\left(\xi\right)}{\left|\xi\right|},
     \,\,\,\,
     {\cal D} \left(T_c\right)= \int_{0}^{\omega^*}d\xi{\cal N}\left(\xi\right)\frac{\tanh\frac{\xi}{2T_c}}{\xi},
\end{equation}
and ${\cal N}\left(\xi\right)$ is the DOS of at a distance $\xi$ away from the Fermi level.
We note that in the case of a four-fold degenerate normal state (spin and valley degenerate phase), the initial $U_0$ is actually $U_0\approx U_{\rm intra}-U_{\rm inter}$, with $U_{\rm intra}$ ($U_{\rm inter}$) being the intra- (inter-) valley interaction strength.
This is a well-known consequence of multiband superconductivity enhanced by the interband interactions.

At a given inverse temperature $\beta=1/T$ the gain in free energy due to condensation of the superconducting phase is given by
\begin{align}
    F_{\Delta}&=\frac{\left|\Delta\right|^{2}}{g_{{\rm eff}.}}-4\int_{0}^{\omega^{*}}d\xi{\cal N}\left(\xi\right)\left(\sqrt{\xi^{2}+\left|\Delta\right|^{2}}-\left|\xi\right|\right)\nonumber\\
    &-\frac{8}{\beta}\int_{0}^{\omega^{*}}d\xi{\cal N}\left(\xi\right)\log\left(\frac{1+e^{-\beta\sqrt{\xi^{2}+\left|\Delta\right|^{2}}}}{1+e^{-\beta\left|\xi\right|}}\right),\label{eq:FDelta}
\end{align}
where the effective coupling constant is precisely the one in Supplementary Equation~\eqref{eq:findTcTAM}, i.e.,
$g_{{\rm eff}.}=g_{\rm att.}-\frac{U_0}{1+U_0\ell}$.
Minimizing $F_\Delta$ with respect to $\Delta$ would produce the analog of $a^2/\left(4b\right)$ from Supplementary Equation~\eqref{eq:GLasymm}.
We note that one may also expand Supplementary Equation~\eqref{eq:FDelta} to find suitable expressions for $a$ and $b$, yet this expansion becomes less reliable for extracting the condensation energy in the regime $T\ll T_c$.

In order to estimate the behavior of the coherence length, we make use of a phenomenological parameter, $\xi_{0}\left(T_{c,0}\right)$.
It is the zero-temperature coherence length expected for a superconductor with critical temperature $T_c$.
We assume here that the relation between the coherence length and the critical temperature remains the same within our regime of interest (although $T_c$ may change considerably).
Employing conventional Ginzburg-Landau relations, we may approximate,
\begin{equation}
    \xi_\Delta \left(T_c,T\right)\approx \xi_{0}\left(T_{c,0}\right)\sqrt{\frac{T_{c,0}^{2}}{T_{c}\left(T_{c}-T\right)}}.
\end{equation}
Therefore, calculation of $T_c$, in combination with the aforementioned phenomenological parameter, will give us $\xi_\Delta$ at any temperature.

\section{Extended data}\label{appsec:extendeddate}

Let us put the first-order transitions we study in the main text in the context of a broader region of the experimental phase diagram, i.e., in a broad regime of total density and displacement field.
In Supplementary Fig.~\ref{fig:phaseBoundaries} we highlight the phase boundaries for the IVC transition in RTG (\ref{fig:phaseBoundaries}a), and for the Stoner transition in BBG (\ref{fig:phaseBoundaries}b), for the same parameters we explore the stability of the false vacuum in.
Additionally, we explore a different region of the RTG phase diagram where a Stoner-like transition may occur (lower density and displacement fields, Supplementary Fig.~\ref{fig:phaseBoundaries}c).
It is clear that on the level of our mean field analysis, the first-order nature of the transition is quite robust over a broad range of variables.

\begin{figure}
    \centering
    \includegraphics[width=18cm]{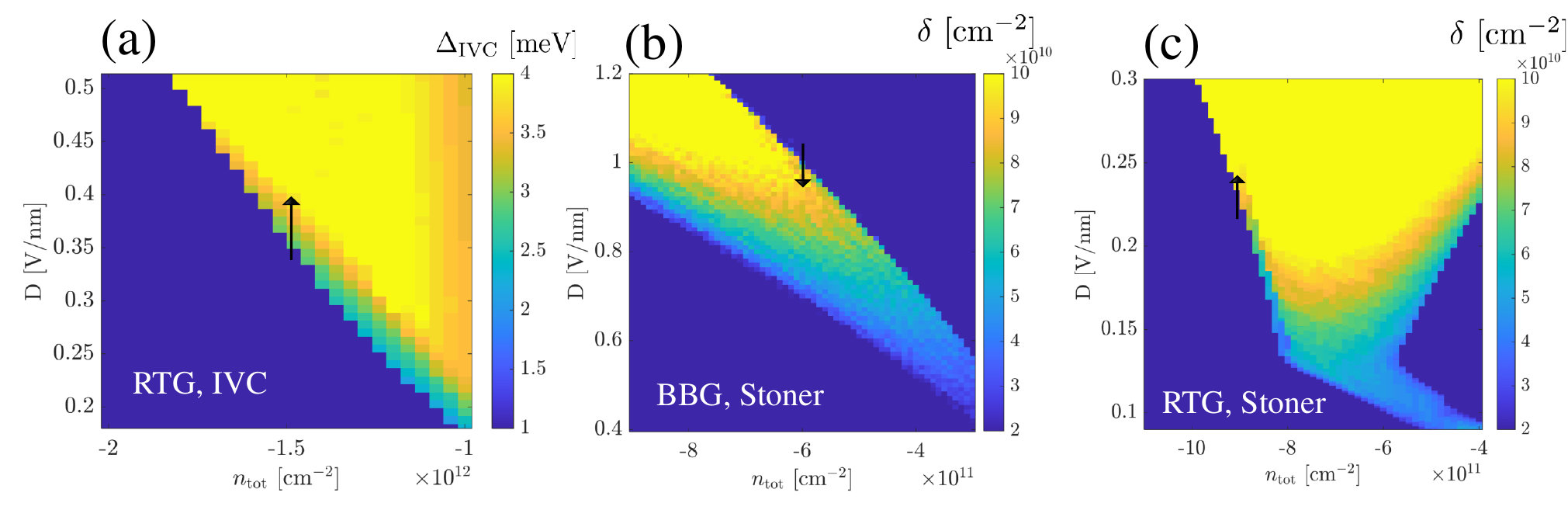}
    \caption{
    Phase boundaries for the symmetry breaking transitions considered in this work.
    Black arrows in each panel mark the transition from a symmetric phase to a symmetry-broken phase at the density considered in the main text or in Supplementary Fig.~\ref{fig:megaDataExtended}.
    The system and the kind of symmetry-breaking which is considered is indicated in the bottom left of each panel.
    (a)
    Self-consistent IVC order parameter in RTG, with interaction parameter $U_C=1.46$ eV nm$^2$.
    (b)
    Self-consistent fourfold degenerate to twofold degenerate polarization in BBG, with interaction parameter $U_C=1.8$ eV nm$^2$.
    (c)
    Self-consistent fourfold degenerate to twofold degenerate polarization in RTG [in a different region than (a)], with interaction parameter $U_C=2.1$ eV nm$^2$.
    }
    \label{fig:phaseBoundaries}
\end{figure}

As stated in the main text, we do not specify how superconductivity is suppressed by the spontaneously formed order.
Here, however, we elaborate on such a possible mechanism near an IVC transition in RTG.
The mechanism can be most readily understood by examination of the DOS near the Fermi level, before or right after the IVC order condenses.
As shown in Supplementary Fig.~\ref{fig:rtgSC_vs_IVC}a, the DOS right at the Fermi energy remains mostly unchanged, so a dramatic effect would seem peculiar.
Moreover, it appears that a large portion of the DOS has actually moved closer to the Fermi level, which should naively only help superconductivity.

The answer to this puzzle is encoded in the calculation of $\ell$, which determines how the initial Coulomb repulsion is screened and renormalized in the Cooper channel.
The renormalization $\ell$ is solely determined by the DOS outside the $\omega^*$ window around the Fermi level, where the retarded attraction becomes effective.
Right outside this retardation window, the prominent DOS feature in the IVC phase is a substantial ``pseudogap'' present between the IVC-split bands.
Thus, following the IVC transition the effective interaction
$g_{{\rm eff}.}=g_{\rm att.}-\frac{U_0}{1+U_0\ell}$
plummets, making the superconductor phase virtually undetectable, see Supplementary Fig.~\ref{fig:rtgSC_vs_IVC}b.

To provide a more complete picture of the quench across the first order phase transition, we demonstrate that such a quench is also possible by varying the total density. This is shown in Supplementary Fig.~\ref{fig:revision_ivc_of_n}.

\begin{figure}
    \centering
    \includegraphics[width=9cm]{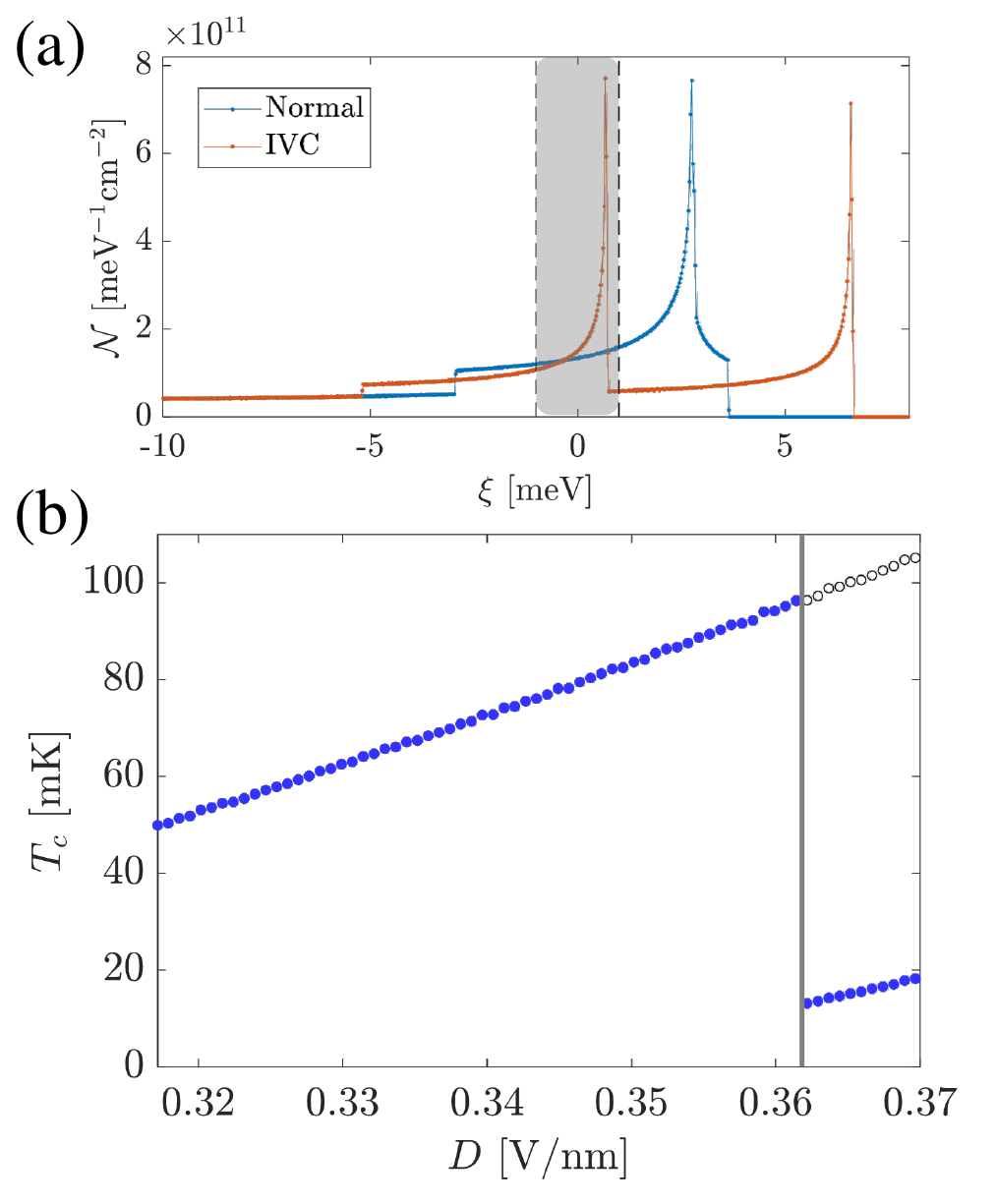}
    \caption{
    (a)
    Density of states as a function of distance from the Fermi level at total density $n_{\rm tot}=-1.5\,10^{-12}$ cm$^{-2}$ for the normal paramagnetic phase (blue), and the IVC phase with $\Delta_{\rm IVC}=2.935$ meV.
    (This value was extracted by minimization of the free energy with interaction strength $g_{\rm IVC}=1.45\, 10^{-11}$ eV nm$^2$.)
    The gray shaded region is the excluded region of width $2\omega^*$, which does not contribute to the repulsion renormalization $\ell$.
    (b)
    Superconducting transition temperature calculated across the IVC transition (blue dots), and without the transition (black circles).
    When the system enters the IVC phase, crossing the vertical gray border, $T_c$ drops below the experimentally detectable $30-40$ mK.
    For the calculations we use $\omega^*=1$ meV, $U_0=0.3$ eV nm$^2$, and $g_{\rm att.}=0.118$ eV nm$^2$.
    }
    \label{fig:rtgSC_vs_IVC}
\end{figure}

\begin{figure}
    \includegraphics[width=11cm]{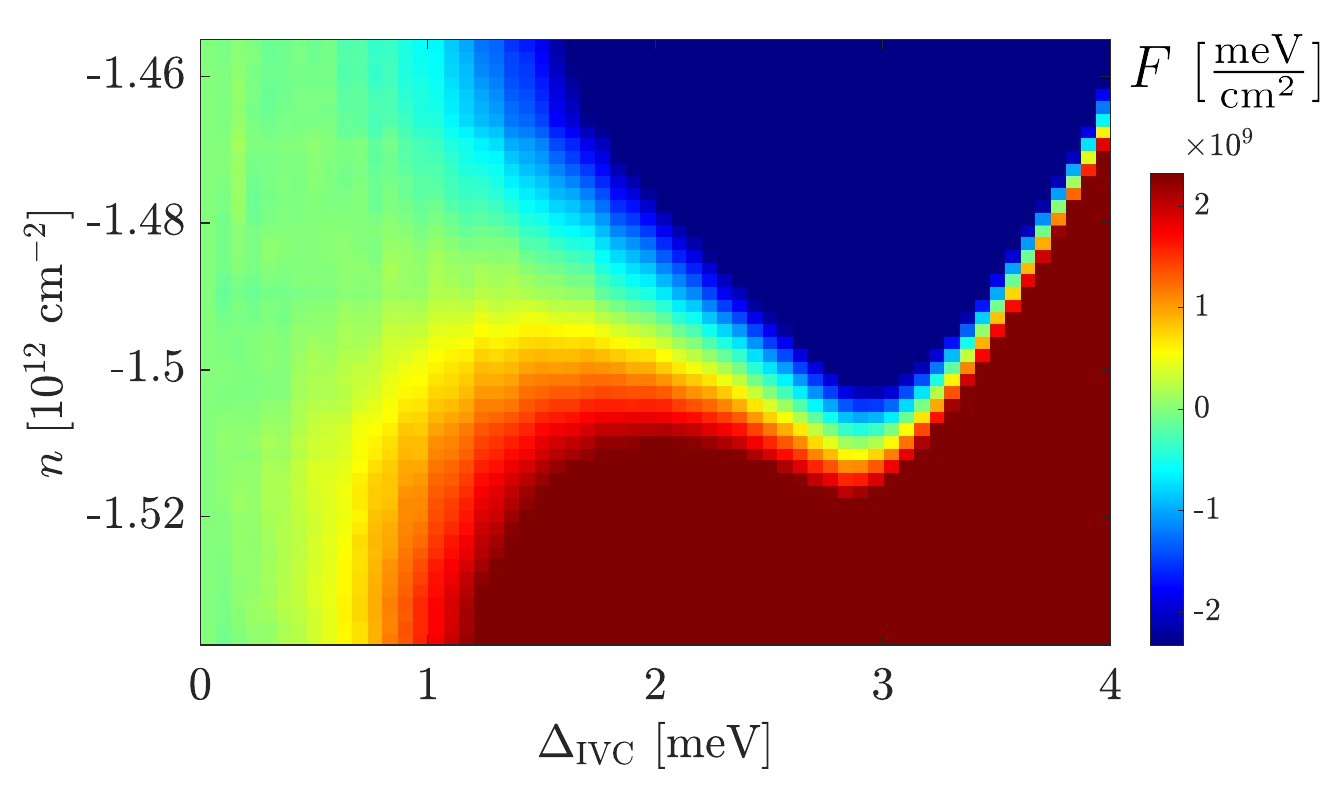}
    \centering
    \caption{{Quenching the density instead of displacement field.} 
    Free energy landscape for the IVC phase transition in RTG as a function of total density.
    A metastable regime is visible.
    We use $U_C=1.46$ eV nm$^2$, as in the main text, and the displacement field $D=0.36$ V/nm.
    }
\label{fig:revision_ivc_of_n}
\end{figure}

Moreover, for the transitions explored in the main text, as well as the RTG Stoner transition, we provide some extended results regarding the metastability diagrams (left column of Supplementary Fig.~\ref{fig:megaDataExtended}), the superconducting parameters needed for extracting the energetics relevant to the superconducting phase (Supplementary Fig.~\ref{fig:megaDataExtended}, center column), and the size of the free-energy barriers and the role of the secondary superconducting order in determining them (right column of Supplementary Fig.~\ref{fig:megaDataExtended}).
Qualitatively, the results share great similarities.
This suggests that the sort of physical phenomena we discuss in this work is quite robust, and should appear quite ubiquitously in theses strongly correlated graphene systems.

\begin{figure}
    \centering
    \includegraphics[width=18cm]{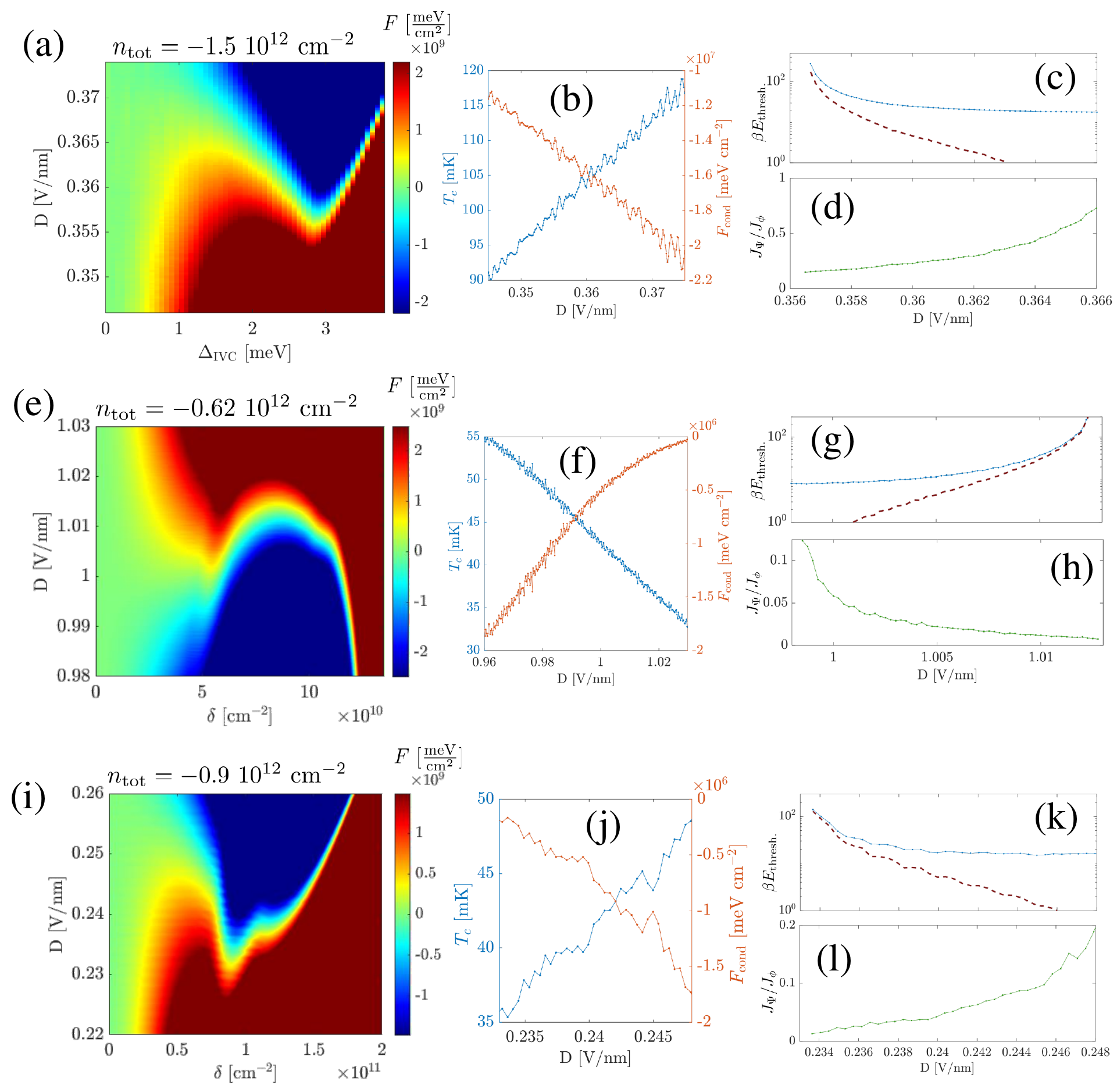}
    \caption{
    Extended results of the considered quenched first-order phase transitions.
    (a)
    Metastability phase diagram of the IVC phase in RTG at a fixed density (indicated above).
    Inset: horizontal cut corresponding to the dashed black line in the main panel.
    The left and right circles are schematics of the Fermi surface in the normal and IVC phases, respectively.
    (b)
    Superconducting $T_c$ (left, blue) and superconducting condensation energy density (right, red) in the vicinity of the transition.
    (c)
    The decay threshold energy (blue) in the vicinity of the transition.
    The dashed purple line corresponds to the result in the absence of the superconducting phase.
    (d) 
    Comparison of the surface tension contributions from the superconducting and IVC phases.
    The parameters used: $T=30$ mK $U_C=1.46$ eV nm$^2$, $g_{\rm att.}=0.21$ eV nm$^2$, $\omega^*=0.5$ meV.
    (e)--(h)
    Same as (a)--(d), for the Stoner transition in BBG.
    The parameters used: $T=30$ mK, $U_C=1.8$ eV nm$^2$, $g_{\rm att.}=0.265$ eV nm$^2$, $\omega^*=0.5$ meV.
    (i)--(l)
    Same as (a)--(d), for the Stoner transition in RTG (not discussed in main text, but appears in Supplementary Fig.~\ref{fig:phaseBoundaries}c).
    The parameters used: $T=30$ mK, $U_C=2.1$ eV nm$^2$, $g_{\rm att.}=0.17$ eV nm$^2$, $\omega^*=0.5$ meV.
    }
    \label{fig:megaDataExtended}
\end{figure}

Finally, we provide the relevant critical droplet radius estimates, $R_c$, for the three different transitions considered above.
These are presented in Supplementary Fig.~\ref{fig:revision_Rc}.

\begin{figure}
    \includegraphics[width=17cm]{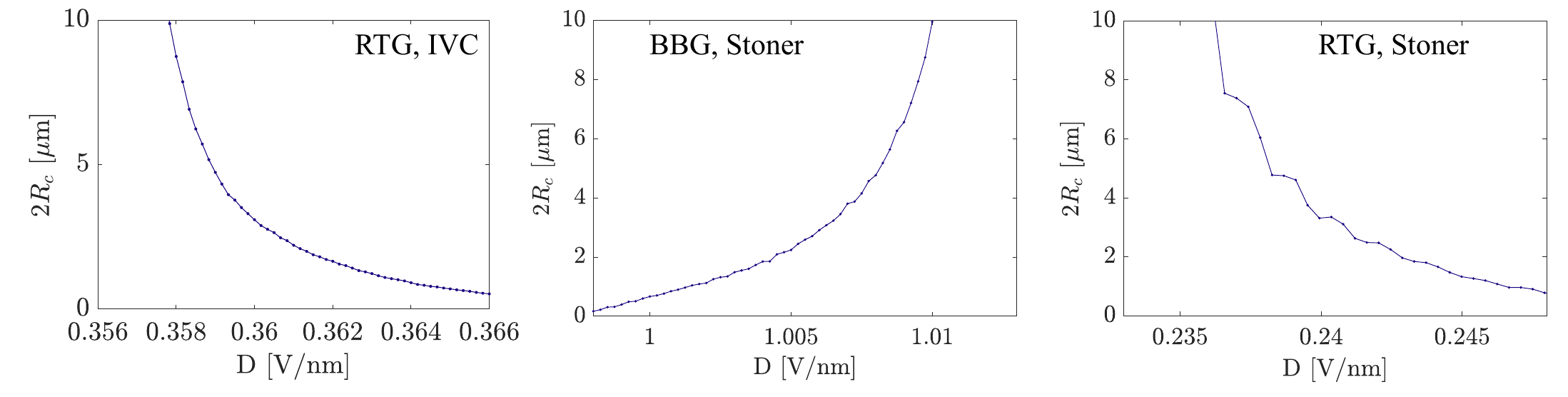}
    \centering
    \caption{{Estimated diameter of the critical droplet for the three examples presented Supplementary Figure~\ref{fig:megaDataExtended}.} 
    Each panel is labeled according to the corresponding system and first-order transition we study.
    Characteristic critical droplet sizes are of the order of microns, comparable to relevant device dimensions.
    The parameters used in extraction of $R_c$ are identical to the ones used for the presented results in Supplementary~\ref{fig:megaDataExtended}.
    }
\label{fig:revision_Rc}
\end{figure}

\section{Heuristic master equation}\label{appsec:masterequationHeuristic}
Let us consider a heuristic approach to the state of the system, where three points of interest exist in phase space.
The first, is the true ground state of the system, to which we assign the occupation probability $p_{\rm true}$.
The second, is the false vacuum state, with $p_{\rm false}$.
Lastly we have also the normal phase, where both of the phases, i.e., $\hat{\phi}$ and $\hat{\Psi}$, are disordered, to which we assign $p_0$.
We may estimate the state of the system as a function of time by considering the time evolution
\begin{equation}
    \frac{d}{dt}\begin{pmatrix}p_{{\rm true}}\\
p_{0}\\
p_{{\rm false}}
\end{pmatrix}=\begin{pmatrix}0 & \gamma_{0} & \gamma\\
0 & -\gamma_{\hat{\Psi}}-\gamma_{0} & 0\\
0 & \gamma_{\hat{\Psi}} & -\gamma
\end{pmatrix}\begin{pmatrix}p_{{\rm true}}\\
p_{0}\\
p_{{\rm false}}
\end{pmatrix}.
\end{equation}
Here, $\gamma$ corresponds to $\tau_{\rm decay}^{-1}$ which we compute in the presence of the secondary competing order (so one always expects $\gamma_0>\gamma$), $\gamma_0$ is the decay rate in the absence of this competition, and $\gamma_{\hat{\Psi}}$ is the time it takes the $\hat{\Psi}$ phase to form in the system starting from the normal state (assuming the absence of the true vacuum order).

Essentially, there are two scenarios of interest, depending on the initial conditions at the time of the quench. 
In the first case, the system starts in the false vacuum state,
$\left(p_{\rm true}\left(0\right),p_0\left(0\right),p_{\rm false}\left(0\right)\right)=\left(0,0,1\right)$.
This is the scenario relevant for example in RTG next to the supposed IVC transition, when the system is quenched from the superconducting state into the IVC regime.
In that case, $p_0$ becomes irrelevant and we recover the expected decay,
\begin{equation}
    p_{\rm false}\left(t\right) = e^{-\gamma t} = e^{-t/\tau_{\rm decay}}.
\end{equation}

In the second more complicated case, the initial conditions are
$\left(p_{\rm true}\left(0\right),p_0\left(0\right),p_{\rm false}\left(0\right)\right)=\left(0,1,0\right)$.
The system is quenched from a normal phase into a regime where formation of the $\hat{\Psi}$ order competes with the false vacuum.
The solution for $p_{\rm false}$ is less simple yet quite straightforward,
\begin{equation}
    p_{{\rm false}}\left(t\right)=\frac{\gamma_{\hat{\Psi}}}{\gamma_{\hat{\Psi}}+\gamma_{0}-\gamma}\left[1-e^{-\left(\gamma_{\hat{\Psi}}+\gamma_{0}-\gamma\right)t}\right]e^{-\gamma t}.
\end{equation}

In the interesting scenario, where the superconducting-like phase is formed first and $\gamma_{\hat{\Psi}}$ is the largest rate, the false vacuum first forms in a time-scale $\sim\gamma_{\hat{\Psi}}^{-1}$, occupies a fraction determined by the ratio of the different rates, and then goes on to decay with the characteristic time-scale $\tau_{\rm decay}$ as before.

\section{Order parameter relaxation}\label{appsec:OPrelaxationSC}
Let us estimate the manner by which the superconducting order parameter relaxation would manifest in measurements of $t_d$, the delay time between a current driven through the system, and the appearance of a voltage signal.
Our starting point is the time-dependent Ginzburg-Landau equation~\cite{larkin2005theory} (neglecting the role of the sub-leading thermodynamic order parameter fluctuations),
\begin{equation}
    -\frac{a}{2}\tau_{GL}\left(\partial_{t}+2ie\varphi\right)\Psi=\frac{\delta F}{\delta\Psi^{*}}.
\end{equation}
Here, $F$ is the functional introduced in Eq.~\eqref{eq:GLasymm} (disregarding the coupling to the $\hat{\phi}$-sector, to be considered shortly), and $\varphi$ is the scalar potential.
We rewrite the equation in terms of the normalized order parameter $\Psi\equiv\sqrt{\frac{b}{a}}\psi$ ($a>0$ in the ordered phase), and find
\begin{equation}
\tau_{GL}\left(\partial_{t}+2ie\varphi\right)\psi=2\xi_{\Psi}^{2}\partial_{x}^{2}\psi+\psi\left(1-\left|\psi\right|^{2}\right).\label{eq:TDGLuseful}
\end{equation}
Notice we assumed the possibility of current flowing solely in the $\hat{x}$ direction with density
\begin{equation}
    j_{s}=\frac{ie}{m^{*}}\frac{a}{b}\left(\psi^{*}\partial_{x}\psi-\psi\partial_{x}\psi^{*}\right),\label{eq:js}
\end{equation}
and $m^*$ is the effective mass.

We employ the ansatz $\psi\left(x\right)=f e^{i\theta\left(x\right)}$, assuming that $f$ remains fairly constants over length scales $\gg\xi_{\Psi}$.
The current may then be expressed as a function of $\partial_x \theta$,
$j_s = -\frac{2e}{m}\frac{a}{b}\partial_{x}\theta\left[1-2\xi_{\Psi}^{2}\left(\partial_{x}\theta\right)^{2}\right]$, and maximizing over this variable allows one to obtain the critical current
\begin{equation}
    j_{c}=-\frac{e}{m^*}\frac{a}{b}\frac{4}{3\sqrt{6}\xi_{\Psi}}.
\end{equation}

We now combine our ansatz, the definition of the current operator, Eq.~\eqref{eq:js}, and the real part of Eq.~\eqref{eq:TDGLuseful}, and find the following connection between the current density and the relaxation of the order parameter,
\begin{equation}
    \tau_{GL}\partial_{t}f=-\left(\frac{j_{s}}{j_{c}}\right)^{2}\frac{4}{27f^{3}}+f\left(1-f^{2}\right).\label{eq:relaxation_standard}
\end{equation}
In the standard experiment demonstrated in Ref.~\cite{relaxation_SC_PALS1979150}, one obtains the delay time by integrating Eq.~\eqref{eq:relaxation_standard} over the change of $f$ from unity to zero,
\begin{equation}
    \frac{t_d^0}{\tau_{GL}} = \int^1_0 \frac{df}{\left(\frac{I}{I_{c}}\right)^{2}\frac{4}{27f^{3}}-f+f^3},
\end{equation}
where $I$ and $I_c$ are the applied current and critical current.
The delay time $t_d^0$ diverges when approaching $I_c$ from the $I>I_c$ side, and is infinite (by definition) for $I<I_c$.

Next, we modify Eq.~\eqref{eq:relaxation_standard} as to be appropriate for our considered scenario.
As a first order approximation, we introduce this modification via the critical current alone.
We assume an independent temporal decay of the superfluid density $\left|\Psi_0\right|^2=\frac{a}{b}$ with a time scale $\tau_{\rm decay}\gg\tau_{GL}$, facilitating a similar decay of the critical current.
As a result, the modified evolution we consider is
\begin{equation}
    \tau_{GL}\partial_{t}f=-e^{2t/\tau_{{\rm decay}}}\left(\frac{j_{s}}{j_{c}}\right)^{2}\frac{4}{27f^{3}}+f\left(1-f^{2}\right).\label{eq:relaxationFULL}
\end{equation}
The validity of Eq.~\eqref{eq:relaxationFULL} becomes questionable around $j_s\to 0$, as one finds relaxation times much longer than $\tau_{\rm decay}$, violating our working assumption.
This is due to the fact that Eq.~\eqref{eq:relaxationFULL} implicitly assumes the dominance of the GL-induced relaxation over the decay.
Thus, it is valid strictly in the regime
\begin{equation}
    j_{s}\gg j_{c}\sqrt{\frac{\tau_{GL}}{\tau_{\rm decay}}}.
\end{equation}
\end{widetext}

\end{document}